\documentclass [nofootinbib,superscriptaddress]{revtex4}
\usepackage [latin9]{inputenc}
\usepackage{mathtools}
\usepackage{amsmath}
\usepackage{amssymb}
\usepackage{graphicx}
\usepackage{esint}

\makeatletter
\providecommand{\tabularnewline}{\\}
\@ifundefined{textcolor}{}
{%
 \definecolor{BLACK}{gray}{0}
 \definecolor{WHITE}{gray}{1}
 \definecolor{RED}{rgb}{1,0,0}
 \definecolor{GREEN}{rgb}{0,1,0}
 \definecolor{BLUE}{rgb}{0,0,1}
 \definecolor{CYAN}{cmyk}{1,0,0,0}
 \definecolor{MAGENTA}{cmyk}{0,1,0,0}
 \definecolor{YELLOW}{cmyk}{0,0,1,0}}
\usepackage{epsfig}
\usepackage{amsfonts}
\usepackage{color}
\usepackage{graphics}
\usepackage{epstopdf}
\usepackage{adjustbox}
\usepackage{lipsum}
\usepackage{subfloat}
\usepackage{slashed}

\definecolor{blu}{cmyk}{1,0.7,0,0.6}
\makeatother

\begin{document}

\title{\color{blu}Probing radiative neutrino mass models using trilepton
channel at the LHC}

\author{Dounia Cherigui}

\email{dounia.cherigui@univ-usto.dz}

\author{Chahrazed Guella}

\email{chahra.guella@gmail.com}

\affiliation{Faculté de Physique, Département de Génie Physique, Université des
sciences et de la technologie, BP 1505, Oran, El M'Naouer, Algerie}

\author{Amine Ahriche}

\email{aahriche@ictp.it}

\affiliation{Department of Physics, University of Jijel, PB 98 Ouled Aissa, DZ-18000
Jijel, Algeria}

\affiliation{The Abdus Salam International Centre for Theoretical Physics, Strada
Costiera 11, I-34014, Trieste, Italy.}

\affiliation{Department of Physics and Center for Theoretical Sciences, National
Taiwan University, Taipei 106, Taiwan.}

\author{Salah Nasri}

\email{snasri@uaeu.ac.ae}

\affiliation{Department of physics, United Arab Emirates University, P.O. Box
15551, Al-Ain, United Arab Emirates.}

\affiliation{Laboratoire de Physique Théorique, DZ-31000, Es-Senia University.
Oran, Algeria.}
\begin{abstract}
In this work, we probe a class of neutrino mass models through the
lepton flavor violating interactions of a singlet charged scalar,
$S^{\pm}$ at the LHC proton-proton collisions with 8 TeV and 14 TeV
energies. This scalar couples to the leptons and induces many processes
such as $pp\rightarrow\ell^{\pm}\ell^{\pm}\ell^{\mp}+\slashed{E}_{T}$.
In our analysis we discuss the opposite sign same flavor leptons signal,
as well as the background free channel with the tau contribution which
can enhance the signal/background ratio for center of mass energies
$\sqrt{s}$= 8 TeV and $\sqrt{s}$ = 14 TeV. 
\end{abstract}
\maketitle

\section{Introduction}

There are number of motivations why the standard model (SM) of particle
physics needs to be extended with new degrees of freedom. This includes
the observation of neutrino oscillations for which the data can not
be explained by massless neutrinos, the nature of dark matter (DM),
and the origin of the matter-antimatter asymmetry of the universe.

One of the most popular mechanisms that generates small neutrino mass
is the seesaw mechanism which comes in different types: type-I ~\cite{seesaw},
the type-II~\cite{seesaw0,seesawII} and type-III ~\cite{seesawIII}.
This mechanism introduces new particles many orders of magnitude heavier
than the electroweak scale that give rise to tiny neutrino mass after
being integrated out from the low energy theory. To avoid fine tuning
of the SM couplings, the mass scale of the new particles needs to
be of order $10^{12}~\text{\textrm{GeV}}$ which makes the high scale
see-saw mechanism impossible to be test at laboratory experiments.
In addition, for such superheavy mass scale the electroweak vacuum
can be destabilized~\cite{DHT}. Other alternative realizations invoking
`low-scale mechanisms' were proposed in~\cite{Boucenna:2014zba}.

Another attractive way to induce naturally small neutrino mass is
the radiative neutrino mass generation, where neutrino mass are generated
at loop level~\cite{rad,Ma,knt,kanemura,Okada:2016rav}. Moreover,
the scale of new physics is much smaller than in the conventional
see-saw and can be of the same order as the electroweak scale for
the three-loop radiative neutrino mass models. For instance, the KNT
model proposed in~\cite{knt} extends the SM with two singlet charged
scalars, $S_{1,2}$ , and one singlet fermion, $N$, all having masses
around the TeV scale, making it testable at collider experiments.
Different phenomenological aspects of this model, such as the DM relic
density, were investigated in~\cite{Cheung}. However, in order to
match the neutrino mass and mixing with the experimental data without
being in conflict with the bound on the process $\mu\rightarrow e+\gamma$,
three generations of singlet fermions are required~\cite{AN}. Generalization
of the KNT model was proposed in~\cite{knt3} by promoting $S_{2}$
and $N$ to multiplets of the $SU(2)_{L}$ gauge symmetry. In these
models, the use of a discrete symmetry that precludes the tree-level
mass term for neutrinos allows the existence of a DM candidate which
plays a role in the radiative neutrino mass generation and could also
trigger the electroweak symmetry breaking~\cite{knt-SI}.

Most of the neutrino mass motivated models, based either on radiative
or seesaw mechanisms, contain charged scalar(s) whose interactions
induce lepton flavor violating (LFV) processes, and thus their couplings
are subject to severe experimental constraints~\cite{PDG,Adam:2013mnn}.
Probing these interactions is of great importance to identify through
which mechanism neutrino mass is generated, and whether it is a Dirac
or Majorana particle.

There has been many attempts to investigate different consequences
of the new interactions in models motivated by neutrino mass at future
colliders~\cite{Perez:2008zc}. Ref.~\cite{knt-ilc} investigated
the possibility of testing the KNT model through the process $e^{+}+e^{-}\rightarrow e^{-}\mu^{+}+\slashed{E}_{T}$
at the ILC, where it was shown that it could be probed at ILC at center
of mass energies 500 GeV and 1 TeV with and without the use of polarized
beams. Similar study has been carried out for the processes $pp\rightarrow e^{-}e^{+}\left(\mu^{-}\mu^{+},e^{-}\mu^{+}\right)+\slashed{E}_{T}$
through the production of the charged scalar $S^{\pm}$ via the Drell-Yan
process and their decay modes which can give a detectable signal with
two charged leptons and missing energy in the final state. The observation
of an electron (positron) and anti-muon (muon) (the latter presents
the most favorite channel), give us an indication for the signature
of this class of model, where it has been shown that the LHC@14 TeV
with 100 fb$^{-1}$ luminosity can test this model~\cite{Guella}.

In this work, without refering to a specific model of radiatively
induced neutrino mass, we investigate the effect of the charged scalar
$S^{\pm}$ on the trilepton final state ($\ell^{\pm}\ell^{\pm}\ell^{\mp}$)
at the LHC, where the background consists of processes mediated by
the gauge bosons $WZ(W\gamma^{\ast})$~\cite{Tri}. Then we will
propose sets of benchmark points for different charged scalar masses
and couplings which are consistent with LFV constraints and investigate
the signal feasibility within the CMS analysis~\cite{CMS-cuts}.

This paper is organized as follows. In section II, we describe the
model and different experimental constraints. Then in section III,
we use the 8 TeV LHC RUN-I data to put constraints on this class of
models and probe the model at 14 TeV. In section IV, we consider two
benchmark points and perform detailed analysis. Possible test of this
class of models through a LFV background free process is investigated.
Finally, we give our summary.

\section{Model \& Space Parameter}

In this work, we consider a class of models that contain the following
term in the Lagrangian~\cite{rad,knt,knt3,knt-SI,Zee-ref} 
\begin{equation}
\mathcal{L}\supset f_{\alpha\beta}L_{\alpha}^{T}C\epsilon L_{\beta}S^{+}-m_{S}^{2}S^{+}S^{-}+\mathrm{h.c.},\label{LL}
\end{equation}
where $L_{\alpha}$ is the left-handed lepton doublet, C is the charge
conjugation operator, $\epsilon$\ is the anti-symmetric tensor,
$f_{\alpha\beta}$ are Yukawa couplings which are antisymmetric in
the generation indices $\alpha$ and $\beta$, and $S^{\pm}$ is an
$SU(2)_{L}$-singlet charged scalar field. The interactions above
induce LFV processes such as $\mu\rightarrow e+\gamma$ and $\tau\rightarrow\mu+\gamma$,
with branching fractions 
\begin{align}
\mathcal{B}(\mu & \rightarrow e+\gamma)\simeq\frac{\alpha_{\text{em}}\upsilon^{4}}{384\pi}\frac{|f_{\tau e}^{\ast}f_{\mu\tau}|^{2}}{m_{S}^{4}},\label{meg}\\
\mathcal{B}(\tau & \rightarrow\mu+\gamma)\simeq\frac{\alpha_{\text{em}}\upsilon^{4}}{384\pi}\frac{|f_{\tau e}^{\ast}f_{\mu e}|^{2}}{m_{S}^{4}},\label{tmug}
\end{align}
where $\alpha_{\text{em}}$ is the fine structure constant, and $\upsilon=246~\mathrm{GeV}$
is the vacuum expectation value of the neutral component in the SM
scalar doublet field. These two branching ratios must satisfy the
experimental bounds $\mathcal{B}\left(\mu\rightarrow e+\gamma\right)<5.7\times10^{-13}$~\cite{Adam:2013mnn}
and $\mathcal{B}\left(\tau\rightarrow\mu+\gamma\right)<4.8\times10^{-8}$~\cite{PDG}.
Moreover, a new contribution to the muon's anomalous magnetic moment
is induced at one-loop, given by 
\begin{equation}
\delta a_{\mu}\sim\frac{m_{\mu}^{2}}{96\pi^{2}}\frac{|f_{e\mu}|^{2}+|f_{\mu\tau}|^{2}}{m_{S}^{2}}.\label{da}
\end{equation}
The constraints on the LFV processes (\ref{meg}), (\ref{tmug}) and
(\ref{da}), implies that $|f_{\alpha\beta}|\lesssim\varsigma m_{S}$,
with $\varsigma$ is a dimensionful constant that depends on the experimental
bounds. This means that the couplings $f$ are suppressed for small
values of the charged scalar mass.

\begin{figure} [h]
\begin{centering}
\includegraphics [width=0.5\textwidth]{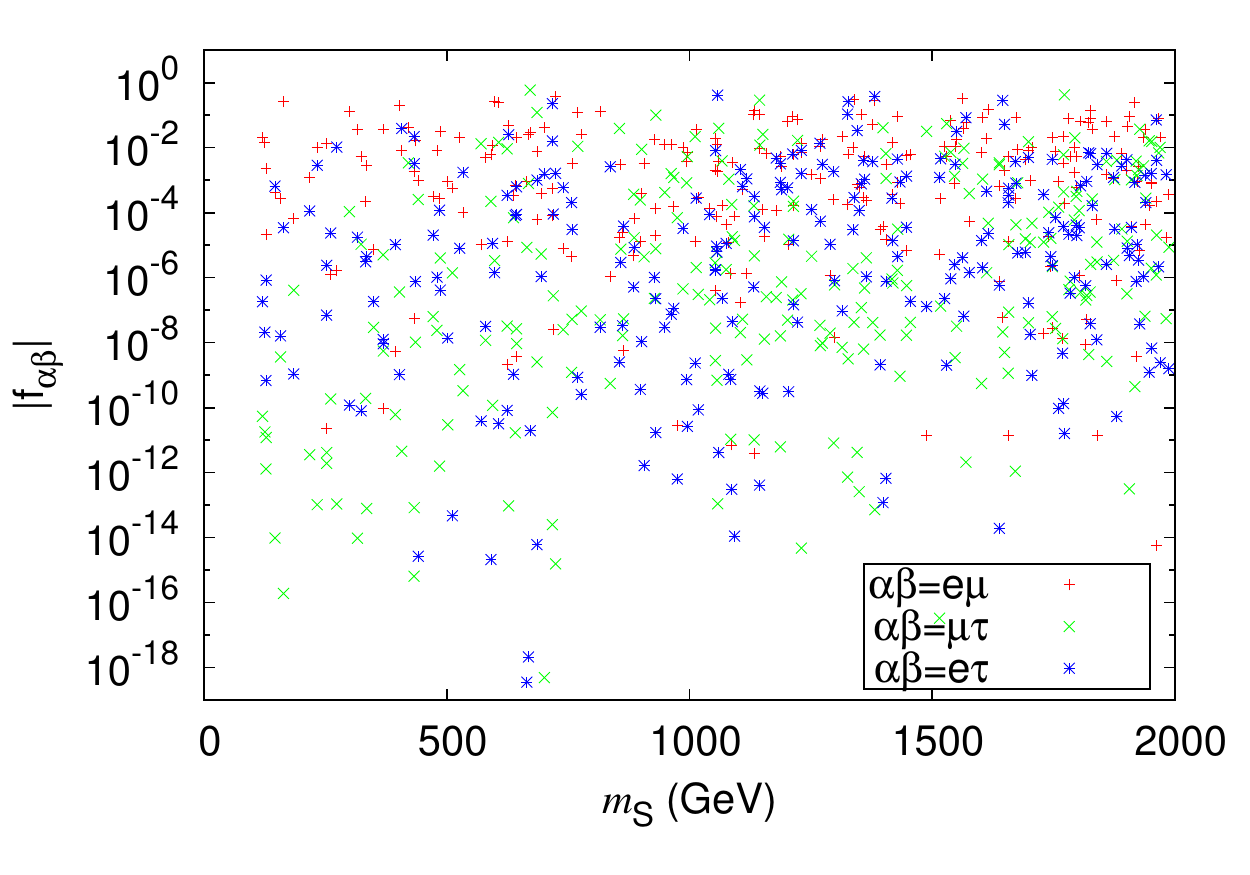}~\includegraphics [width=0.5\textwidth]{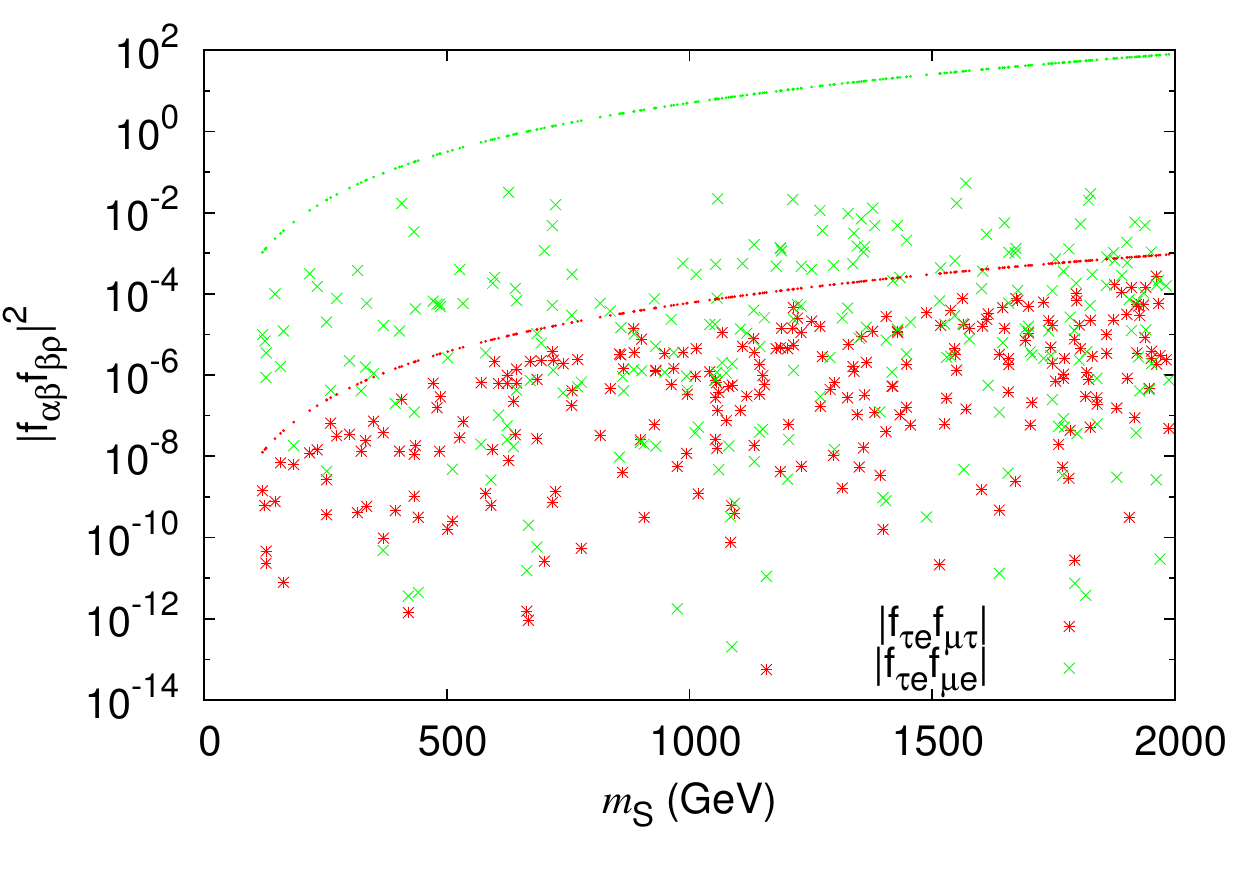} 
\par\end{centering}

\caption{The magnitude of $f$'s versus $m_{S}$ (left) and combination of
the couplings versus $m_{S}$ (right) with the experimental bounds
$\mu\rightarrow e+\gamma$\ and $\tau\rightarrow\mu+\gamma$ are
represented by dashed lines.}
\label{fab} 
\end{figure}

Here, we consider the charged scalar mass in the range $100~\mathrm{GeV}<m_{S}<2~\mathrm{TeV}$,
while the couplings $f_{\alpha\beta}$ take random values that respect
the above mentioned constraints (\ref{meg}), (\ref{tmug}) and (\ref{da}).
These values are illustrated in Fig.~\ref{fab}, where we show the
allowed space parameter for the charged scalar mass and couplings.
It is worth mentioning that the couplings $f_{\alpha\beta}$ shown
in Fig. \ref{fab} could match the observed neutrino oscillation values,
and their values depend on the details of the models ~\cite{AN,knt3,knt-SI,Zee-ref}\footnote{For example, the benchmark points values shown in Table \ref{tab-point}
correspond to the model studied in~\cite{AN}, where the other model
parameters (the couplings $g_{i\alpha}$ and the mass of the other
charged scalar mass) are chosen in a way to match neutrino oscillation
data, DM relic density and LFV constraints. For the models proposed
in~\cite{knt3,knt-SI}, one can adjust the parameters so that most
of the benchmark points shown in Fig. \ref{fab} fulfill the aforementioned
constraints.}. Since our analysis is not restricted to a particular radiatively
induced neutrino mass model, we present a scatter plot in Fig. \ref{fab}-right
for the combination $\left\vert f_{\alpha\rho}f_{\beta\rho}\right\vert ^{2}$
which enter the expressions of the LFV observables. The large overlap
between the region populated by the blue points in Fig.~\ref{fab}-left
plot with the green is due to the fact that they get a common tau
contribution in the expressions of the branching ratios in (\ref{meg})
and (\ref{tmug}), whereas the red points correspond to larger values
of $f_{e\mu}$ as compared to the two other combinations. Fig.~\ref{fab}-right
shows the parameter space region for which upper experimental bounds
of $\mathcal{B}\left(\mu\rightarrow e+\gamma\right)$ and $\mathcal{B}\left(\tau\rightarrow\mu+\gamma\right)$
are satisfied along the identified range of mass noting that this
LFV bounds processes prompt $m_{S}$ to large values once the corresponding
coupling product $f_{\alpha\rho}f_{\beta\rho}$ becomes important.

\section{Current Constraints on Trilepton Signal at the LHC}

\begin{figure} [h]
\begin{centering}
\includegraphics [width=0.9\textwidth]{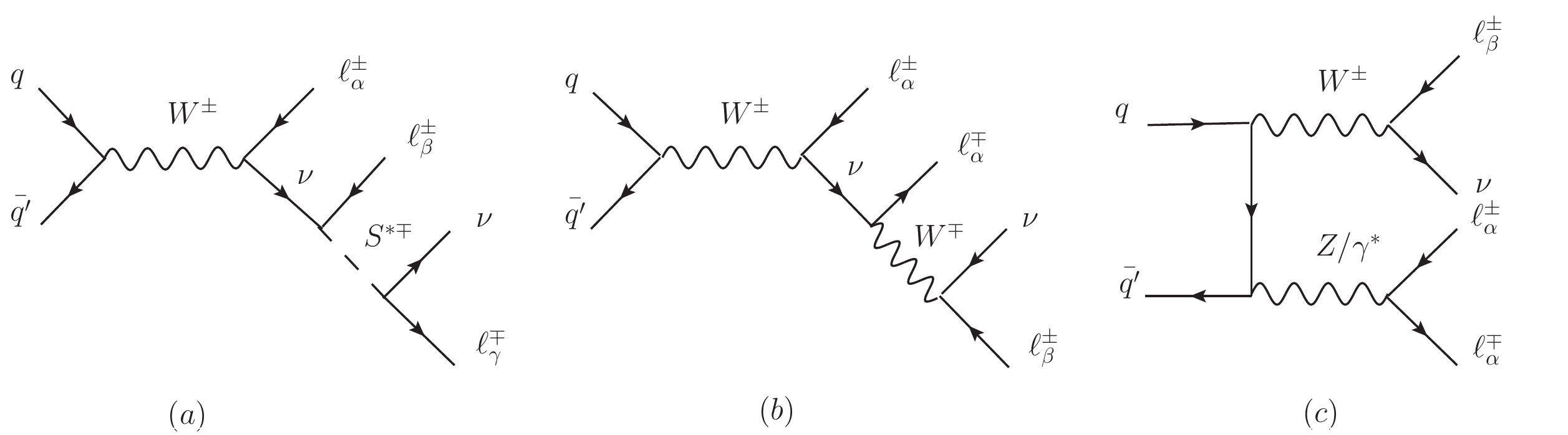} 
\par\end{centering}
\caption{Diagrams corresponding to the trilepton signal (a) and SM background
(b,c).}
\label{tri-p} 
\end{figure}

At the LHC, it is possible to produce a singly charged scalar associated
with different sign different flavor charged leptons through W-boson
exchange which at the parton level read as 
\begin{equation}
q\bar{q}^{\prime}\rightarrow W^{\pm}\rightarrow\ell^{\pm}\ell^{\pm}S^{\ast\mp}\rightarrow\ell^{\pm}\ell^{\pm}\ell^{\mp}+\slashed{E}_{T},
\end{equation}
where the charged scalar $S^{\pm}$ decays into charged lepton and
neutrino giving rise to three leptons plus missing energy in the final
state as shown in Fig.~\ref{tri-p}-a.

According to the diagram presented in Fig.~\ref{tri-p}-a, we have
7 contributions to this trilepton signal: 
\begin{equation}
\ell\ell\ell\equiv ee\mu,e\mu\mu,ee\tau,e\tau\tau,\mu\mu\tau,\mu\tau\tau,e\mu\tau,\label{LLL}
\end{equation}
Here, the process that maximally violates the lepton flavor ($e\mu\tau$)
has a small background, while the other six are accompanied by a large
SM background. Such process with maximal LFV ($e\mu\tau$) can be
a direct probe to the interactions in (\ref{LL}). However, this process
involves purely $S^{\pm}$\-mediated diagrams, and therefore has
a very small cross section due to the smallness of the couplings $f_{\alpha\beta}$
and the heaviness of the charged scalars as dictated by the LFV constraints
(\ref{meg}), (\ref{tmug}) and (\ref{da}). For the processes with
the large SM background, such as $pp\rightarrow e^{\pm}e^{\mp}\mu^{\pm}+\slashed{E}_{T}$,
the transverse missing energy receives two contributions $\slashed{E}_{T}\equiv\nu_{\tau}$,
$\nu_{\mu}$. The process with $\slashed{E}_{T}\equiv\nu_{\tau}$
occurs only through purely $S$-mediated diagrams, and therefore has
a suppressed cross section. However, the second process occurs through
$S$-mediated and W/Z/$\gamma$-diagrams, and hence the cross section
can be written as $\sigma_{M}=\sigma_{SM}+\sigma_{S}+\sigma_{interference}$.
Therefore, the expected excess of events number could be either $\sigma_{S}$
and/or $\sigma_{interference}$, where the former could be significant
only when the charged scalar is on-shell. However we found that $\sigma_{S}/\sigma_{interference}<\mathcal{O}(10^{-5})$
for the benchmark points considered in our analysis. This leads us
to confirm that the event number excess comes mainly from the interference
contribution term.

Due to the difficulty in identifying the tau lepton at the LHC, we
consider in our detailed analysis only the final state leptons $\ell=e,\mu$,
where the missing energy $\slashed{E}_{T}$ can be any neutrino or
antineutrino. The main process that contributes to the SM background
for trilepton production is the irreducible background 
\begin{equation}
q\bar{q}^{\prime}\rightarrow W^{\pm}\rightarrow\ell^{\pm}\ell^{\mp}W^{\pm}\rightarrow\ell^{\pm}\ell^{\pm}\ell^{\mp}+\slashed{E}_{T},~q\bar{q}^{\prime}\rightarrow ZW^{\pm}(\gamma^{\ast}W^{\pm})\rightarrow\ell^{\pm}\ell^{\pm}\ell^{\mp}+\slashed{E}_{T},
\end{equation}
as shown in Fig.~\ref{tri-p}-b and -c. We use CalcHEP~\cite{Belyaev:2012qa}
to generate both the SM background events as well as the events from
processes due to the extra interactions in (\ref{LL}) for CM energies
$\sqrt{s}=8$ TeV and $14$ TeV. Here, the considered values of the
$f_{\alpha\beta}$ Yukawa couplings and the charged scalar mass ($m_{S}$)
make the branching ratios $\mathcal{B}(\mu\rightarrow e+\gamma)$
and $\mathcal{B}(\tau\rightarrow\mu+\gamma)$ just below the experimental
bounds.

In our analysis, we look for the event number difference $N_{ex}=N_{M}-N_{BG}$,
where $N_{M}$ is the expected number of events number coming from
both the new interactions and the SM processes, while $N_{BG}$ is
the background event number. Thus, with integrated luminosity $\mathcal{L}_{int}$
, the excess of events is $N_{ex}=\mathcal{L}_{int}\left(\sigma_{M}-\sigma_{BG}\right)$,
and $N_{BG}=\mathcal{L}_{int}\sigma_{BG}$, with $\sigma_{BG}$ and
$\sigma_{M}$ are the total cross sections due to interactions of
the SM interactions and the one in Eq. (\ref{LL}), respectively,
after imposing the selection cuts. Therefore the signal significance
is given by 
\begin{equation}
S=\frac{N_{ex}}{\sqrt{N_{ex}+N_{BG}}}=\frac{N_{ex}}{\sqrt{N_{M}}}.\label{S}
\end{equation}

One has to mention that the largest source of the SM background is
the multi-jets events which can be misidentified as leptons in the
detector. Among the dominant sources that give rise to these fake
leptons we have the semileptonic decays of the charm and the bottom
quark; and the photons conversion~\cite{CMS:2010lua}. In order to
reduce the contamination in the signal region, we require the electron
events to have $p_{T}>15$ GeV and $\left\vert \eta\right\vert <2.5$,
whereas all the muon candidates are required to have $p_{T}>5$ GeV
and $\left\vert \eta\right\vert <2.4$. The hadronic decay of the
tau charged lepton ${\tau}_{had}$ can be discriminated with $p_{T}>15$
GeV and $\left\vert \eta\right\vert <2.1$~\cite{Chatrchyan:2012ya}.
Additional criteria can be applied in order to suppress the SM background
coming from the QCD-multijet production~\cite{CMS:2010bta}.

In~\cite{CMS-cuts}, the CMS collaboration presented a model-independent
search for anomalous production of events with at least three isolated
charged leptons using their data with an integrated luminosity of
19.5 fb$^{-1}$ at $\sqrt{s}$ = 8 TeV LHC. The analysis is based
on the following criteria:

$\bullet$ The presence of at least three isolated leptons (muon,
electron).

$\bullet$ The transverse momentum of muon and electron must satisfy
$p_{T}^{\ell}>10\ \mathrm{GeV}$.

$\bullet$ The pseudo-rapidity of leptons $|\eta^{\ell}|<2.4$.

$\bullet$ The missing transverse energy $\slashed{E}_{T}<50\ \mathrm{GeV}$.

$\bullet$ In order to remove the low-mass Drell-Yan processes as
well as the 'Below-Z' and 'Above-Z' regions coming from background,
the invariant mass of each opposite sign same flavor lepton pair must
be in the range $75~\mathrm{GeV}<M_{\ell^{+}\ell^{-}}<105~\mathrm{GeV}$.

Using these cuts, it has been found that a bound on the heavy-light
neutrino mixing parameter ($|B_{lN}|^{2}$) for heavy neutrino masses
up to 500 GeV can be established. For instance, $\left\vert B_{lN}\right\vert ^{2}<2\times10^{-3}$
has been derived for $m_{N}\sim100$ GeV~\cite{okada}.

\begin{figure} [h]
\begin{centering}
\includegraphics [width=0.5\textwidth]{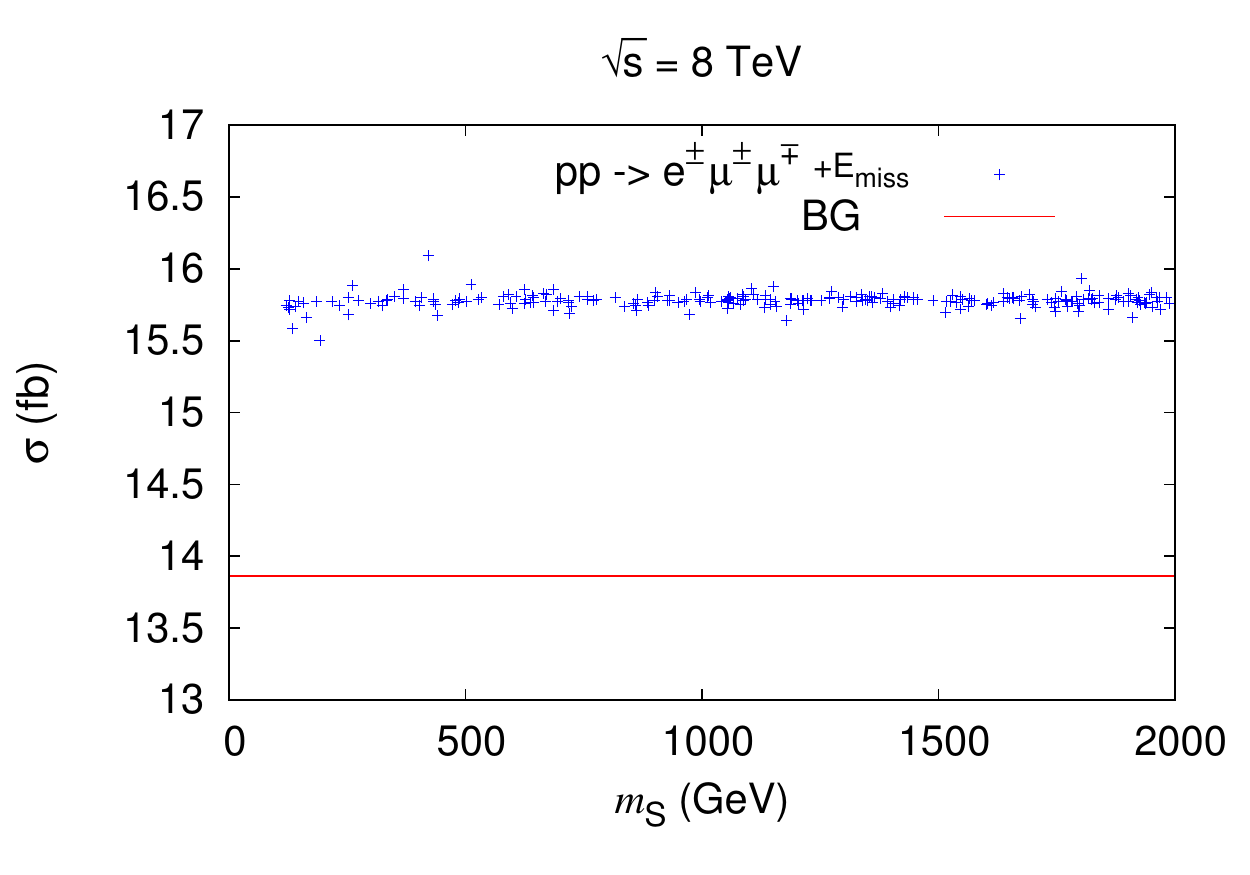}~\includegraphics [width=0.5\textwidth]{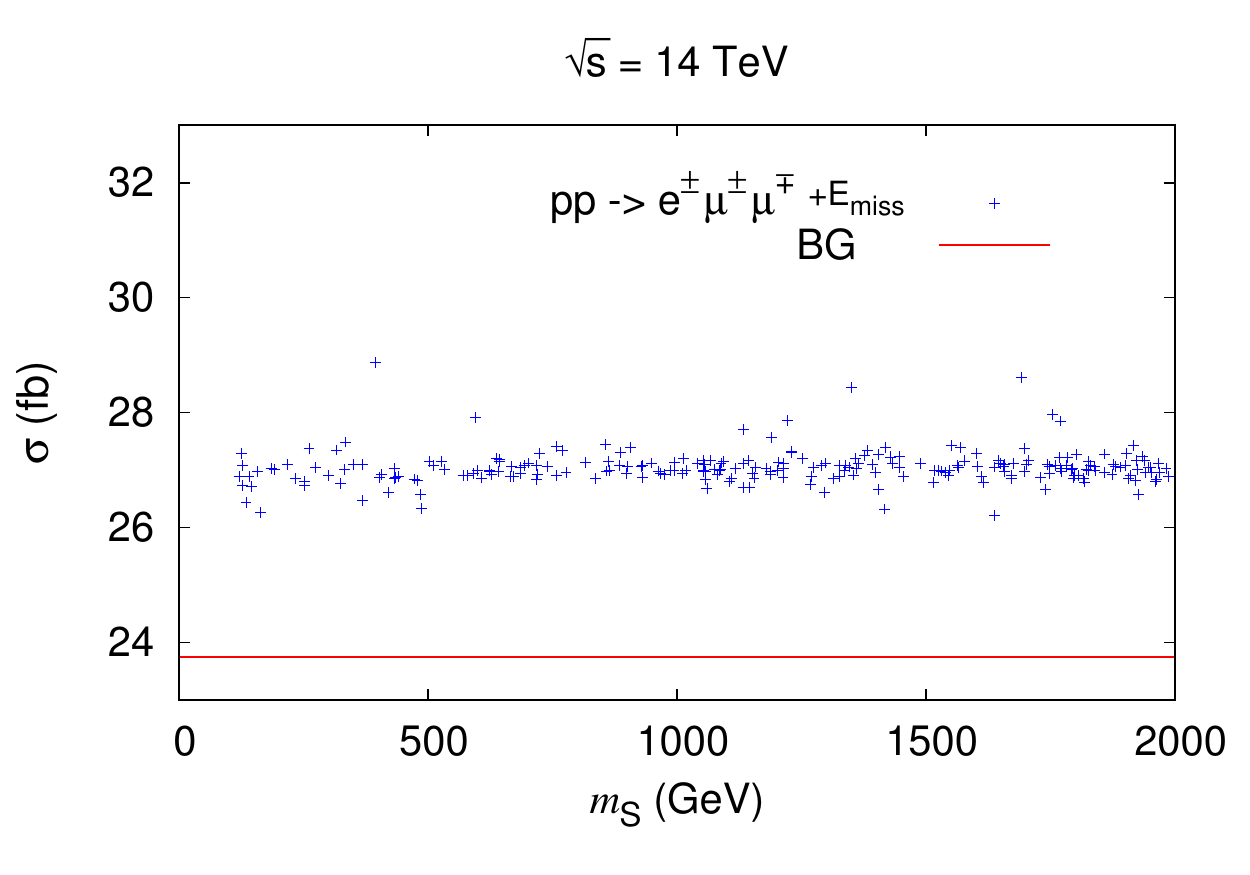}
\includegraphics [width=0.5\textwidth]{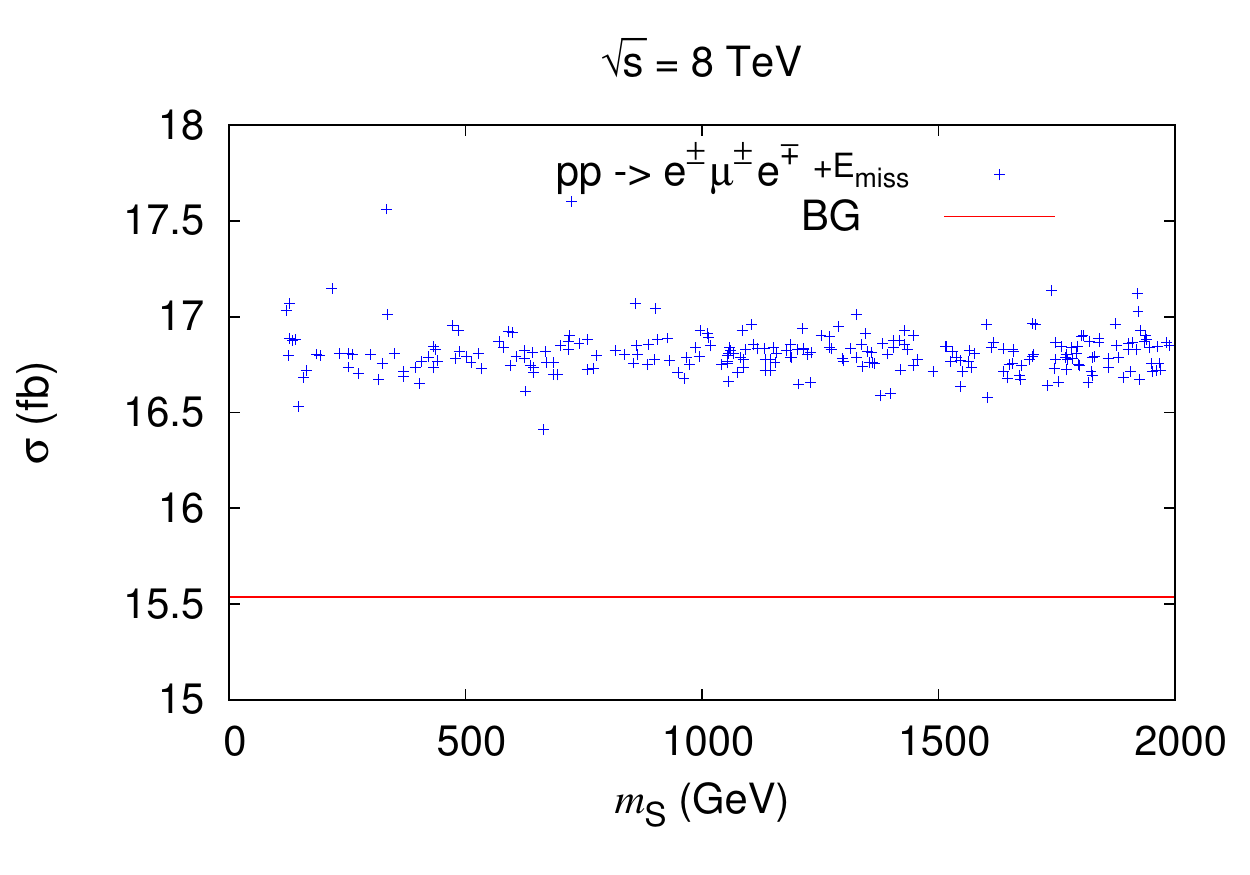}~\includegraphics [width=0.5\textwidth]{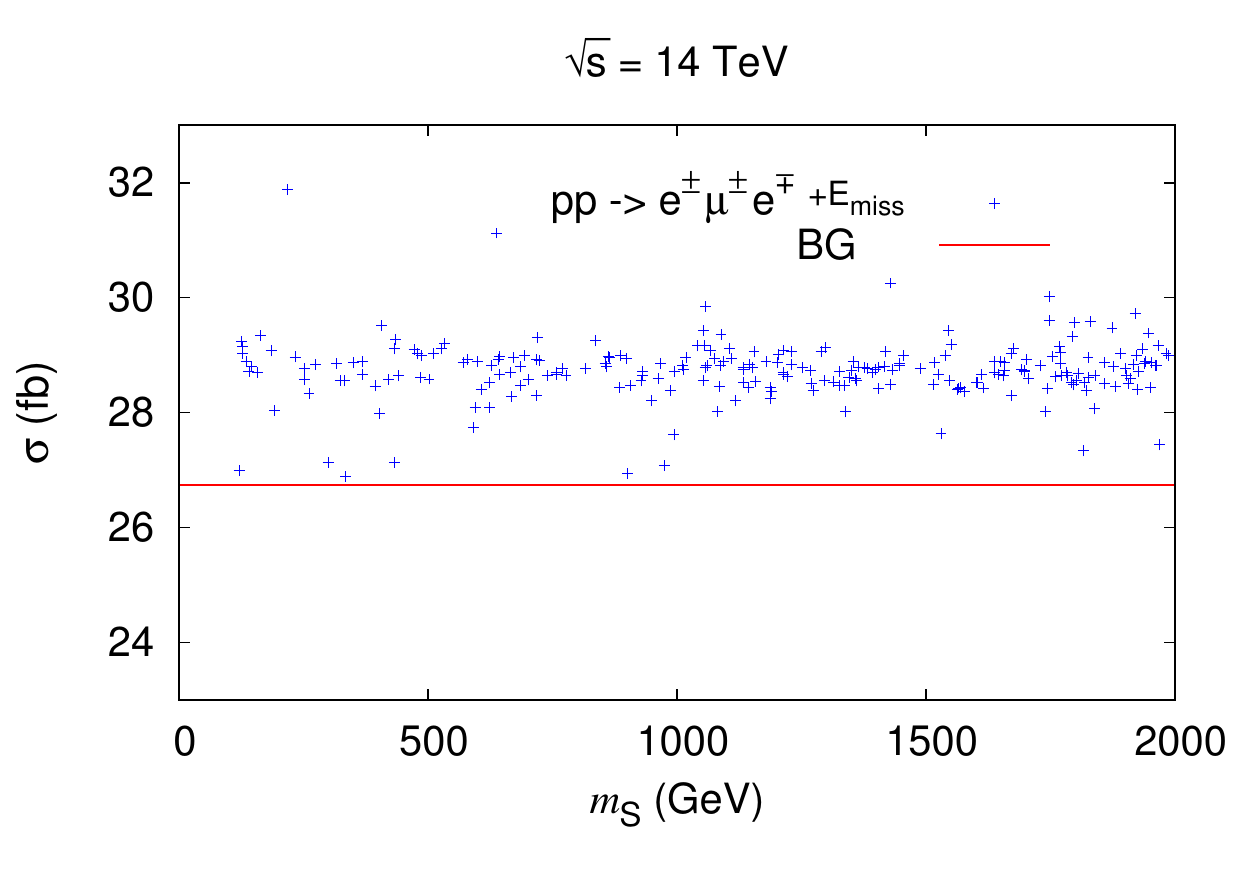} 
\par\end{centering}
\caption{The production cross section for the processes $pp\rightarrow e^{\pm}\mu^{\pm}\mu^{\mp}+\slashed{E}_{T}$
(top), $pp\rightarrow e^{\pm}\mu^{\pm}e^{\mp}+\slashed{E}_{T}$ (bottom)
at $\sqrt{s}=8$ TeV (left) and $\sqrt{s}=14$ TeV (right) as function
of charged scalar mass. The red lines correspond to the background
cross section values.}
\label{CS} 
\end{figure}

In Fig.~\ref{CS}, we show the production cross section $\sigma_{M}$
at the parton level for the first two processes in (\ref{LLL}) as
a function of the charged scalar mass for the benchmark points that
are consistent with experimental bound on the LFV processes discussed
in the previous section.

\begin{figure} [h]
\begin{centering}
\includegraphics [width=0.5\textwidth]{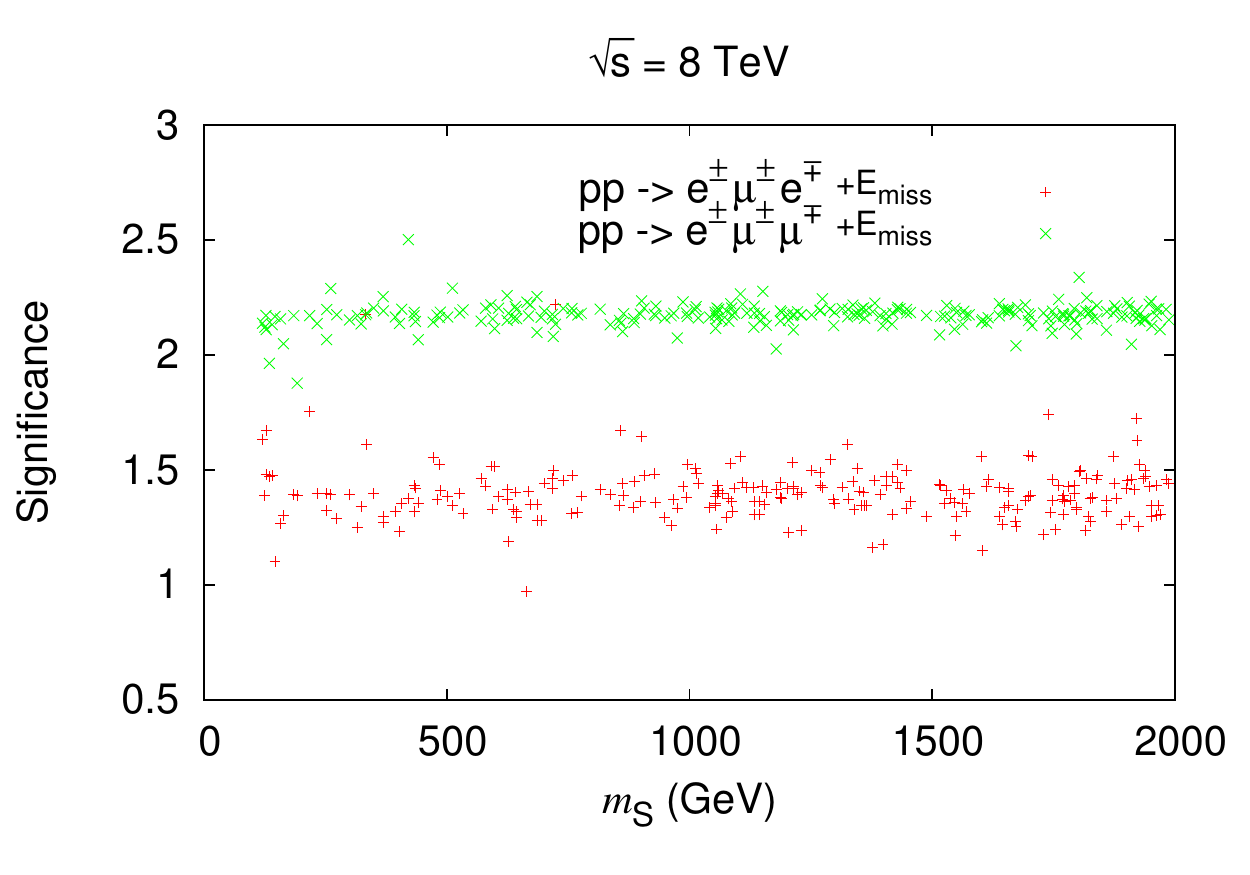}~\includegraphics [width=0.5\textwidth]{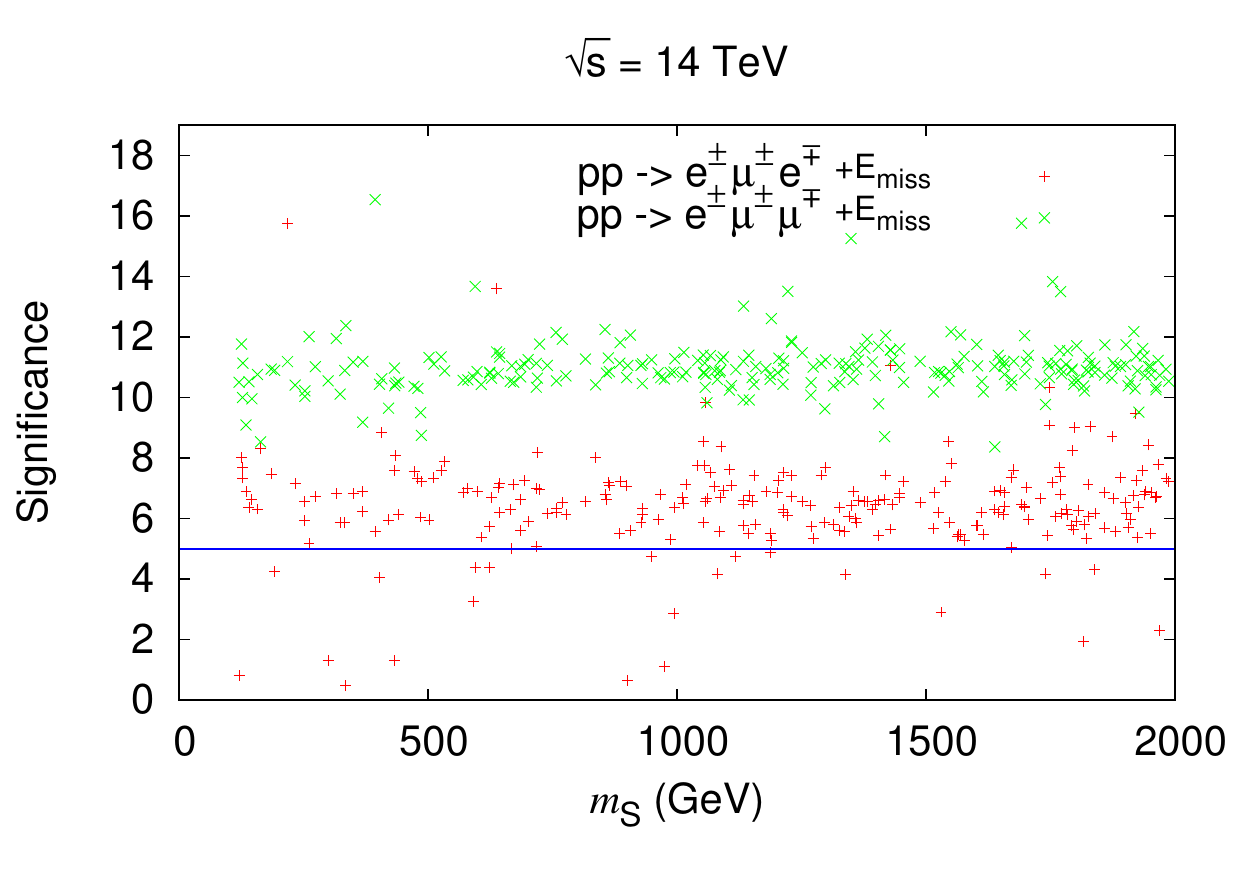} 
\par\end{centering}
\caption{The significance for the process $pp\rightarrow\ell^{\pm}\ell^{\pm}\ell^{\mp}+\slashed{E}_{T}$
at 8 TeV (left) and 14 TeV (right) versus the charged scalar mass
for the integrated luminosity values 20.3 $fb^{-1}$ and 100 $fb^{-1}$,
respectively. The horizontal blue line indicates the significance
value $S=5$.}
\label{sign} 
\end{figure}

We see that $\sigma_{M}$ is larger than the one of the SM background
$\sigma_{BG}$ within the cuts used by the CMS collaboration, and
increases with CM energy whereas it is essentially independent of
the charged scalar mass. To see how important is the signal, we compute
the significance taking into account the previous CMS cuts, for the
two first processes in (\ref{LLL}) for the set of benchmark points
that fulfill the constraints on the LFV processes (\ref{meg}), (\ref{tmug})
and (\ref{da}) that are used previously in Fig.~\ref{fab}. After
applying of the selection criteria quoted above, we show in Fig.~\ref{sign}
the significance for the two considered channels at both 8 TeV and
14 TeV CM energy. 

These results are consistent with searches for new phenomena in events
with multilepton final states, they have not shown any significant
deviation from SM expectations at 8 TeV CM energy. However, after
imposing the same cuts at 14 TeV with 300 fb$^{-1}$ of integrated
luminosity, one shows that it is possible to get at least a 4 sigma
excess for any benchmark point defined in Sec. II. Hence, we carry
this study by searching a significant trilepton signal within this
class of models at $\sqrt{s}$ = 8 TeV and 14 TeV by choosing two
benchmark points and look for different cuts where the significance
could be larger.

\section{Benchmark Analysis}

In this section, we consider two benchmark points, denoted by $B_{1}$
and $B_{2}$, with the charged scalar masses 472 GeV and 1428 GeV
(see Tab.~\ref{tab-point}). Here, we first analyze the trilepton
production with missing energy involving $e$ and $\mu$ decay modes
of the heavy charged scalar $S^{\pm}$ with $\sqrt{s}=8$ TeV and
14 TeV. Then, we discuss possibility of observing the maximally LFV
process signal $\ell^{\pm}\ell^{\pm}\ell^{\mp}\equiv e^{\pm}\mu^{\pm}\tau^{\mp}$.

A critical part in the analysis of signal events associated with new
physics is the accurate estimation of the SM background. For this
purpose, we study the event distributions for the SM background as
well as the background plus the trilepton signal, and impose the cuts
on the relevant observables as shown in Tab.~\ref{tab-cuts}.

\begin{table} [h]
\begin{centering}
\begin{adjustbox}{max width=\textwidth} %
\begin{tabular}{ccccc}
\hline 
Point & $m_{S}$(GeV) & $f_{e\mu}$ & $f_{e\tau}$ & $f_{\mu\tau}$\tabularnewline
\hline 
$B_{1}$ & 472 & -$(9.863+i8.774)\times10^{-2}$ & -$(6.354+i2.162)\times10^{-2}$ & $(0.78+i1.375)\times10^{-2}$\tabularnewline
\hline 
$B_{2}$ & 1428 & $(5.646+i549.32)\times10^{-3}$ & -$(2.265+i1.237)\times10^{-1}$ & -$(0.41-i3.58)\times10^{-2}$\tabularnewline
\hline 
\end{tabular}\end{adjustbox} 
\par\end{centering}

\caption{Two benchmark points selected from the allowed parameter space of
the model.}
\label{tab-point} 
\end{table}

\begin{table} [h]
\begin{centering}
\begin{adjustbox}{max width=\textwidth}%
\begin{tabular}{l|l|l|l|l}
\hline 
\multicolumn{1}{l|}{$e^{\pm}\mu^{\pm}e^{\mp}+\not E_{T}$ @ 8 TeV } & \multicolumn{1}{l|}{$e^{\pm}\mu^{\pm}e^{\mp}+\not E_{T}$ @ 14 TeV } & \multicolumn{1}{l|}{} & \multicolumn{1}{l||}{$e^{\pm}\mu^{\pm}\mu^{\mp}+\not E_{T}$ @ 8 TeV } & $e^{\pm}\mu^{\pm}\mu^{\mp}+\not E_{T}$ @ 14 TeV\tabularnewline
\hline 
\hline 
\multicolumn{1}{c||}{$70<M_{e^{-}e^{+}}<110$} & \multicolumn{1}{c||}{$70<M_{e^{-}e^{+}}<110$} & \multicolumn{1}{c||}{} & \multicolumn{1}{c||}{$80<M_{\mu^{-}\mu^{+}}<100$} & \multicolumn{1}{c}{$80<M_{\mu^{-}\mu^{+}}<110$}\tabularnewline
\multicolumn{1}{c||}{$M_{e^{+}\mu^{+}}<200$} & \multicolumn{1}{c||}{$M_{e^{+}\mu^{+}}<230$} & \multicolumn{1}{c||}{} & \multicolumn{1}{c||}{$M_{e^{+}\mu^{+}}<200$} & \multicolumn{1}{c}{$M_{e^{+}\mu^{+}}<230$}\tabularnewline
\multicolumn{1}{c||}{$M_{e^{-}\nu}<206$} & \multicolumn{1}{c||}{$M_{e^{-}\nu}<220$} & \multicolumn{1}{c||}{} & \multicolumn{1}{c||}{$M_{\mu^{-}\nu}<185$} & \multicolumn{1}{c}{$M_{\mu^{-}\nu}<245$}\tabularnewline
\multicolumn{1}{c||}{$10<p_{T}^{\ell}<100$} & \multicolumn{1}{c||}{$10<p_{T}^{\ell}<90$} & \multicolumn{1}{c||}{} & \multicolumn{1}{c||}{$10<p_{T}^{\ell}<100$} & \multicolumn{1}{c}{$10<p_{T}^{\ell}<130$}\tabularnewline
\multicolumn{1}{c||}{$\left\vert \eta^{\ell}\right\vert <3$} & \multicolumn{1}{c||}{$\left\vert \eta^{\ell}\right\vert <3$} & \multicolumn{1}{c||}{} & \multicolumn{1}{c||}{$\left\vert \eta^{\ell}\right\vert <3$} & \multicolumn{1}{c}{$\left\vert \eta^{\ell}\right\vert <3$}\tabularnewline
\multicolumn{1}{c||}{$\not E_{T}<100$} & \multicolumn{1}{c||}{$\not E_{T}<90$} & \multicolumn{1}{c||}{} & \multicolumn{1}{c||}{$\not E_{T}<120$} & \multicolumn{1}{c}{$\not E_{T}<90$}\tabularnewline
\hline 
\end{tabular}\end{adjustbox} 
\par\end{centering}

\caption{Applied cuts on different kinematical variables: $M_{\ell\ell}$ (invariant
mass), $p_{T}^{\ell}$ (charged lepton transverse momentum), $\protect\not E_{T}$
(transverse missing energy), and $\eta^{\ell}$ (pseudo-rapidity).
The energy dimension variables are in GeV unit.}
\label{tab-cuts} 
\end{table}

\begin{figure} [h]
\begin{centering}
\includegraphics [width=0.33\textwidth]{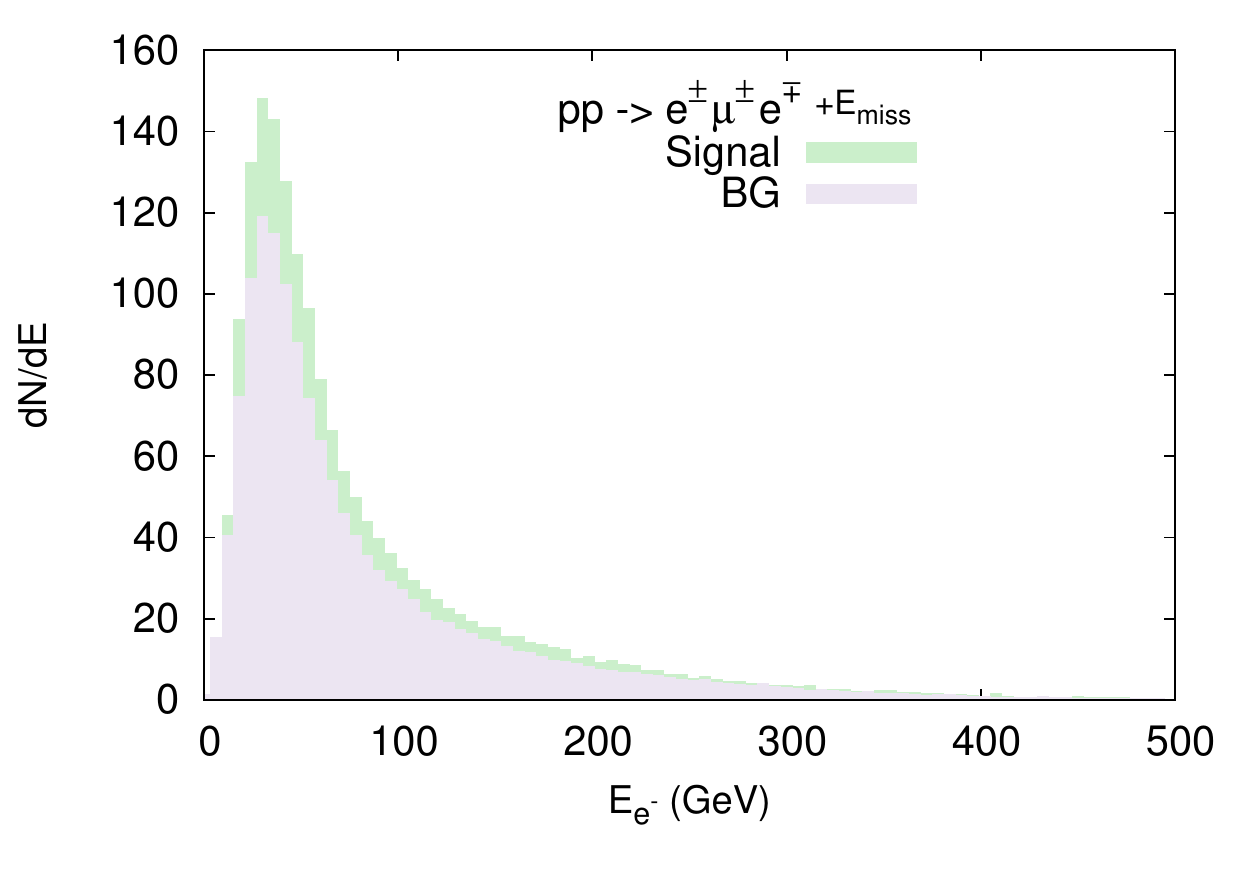}~\includegraphics [width=0.33\textwidth]{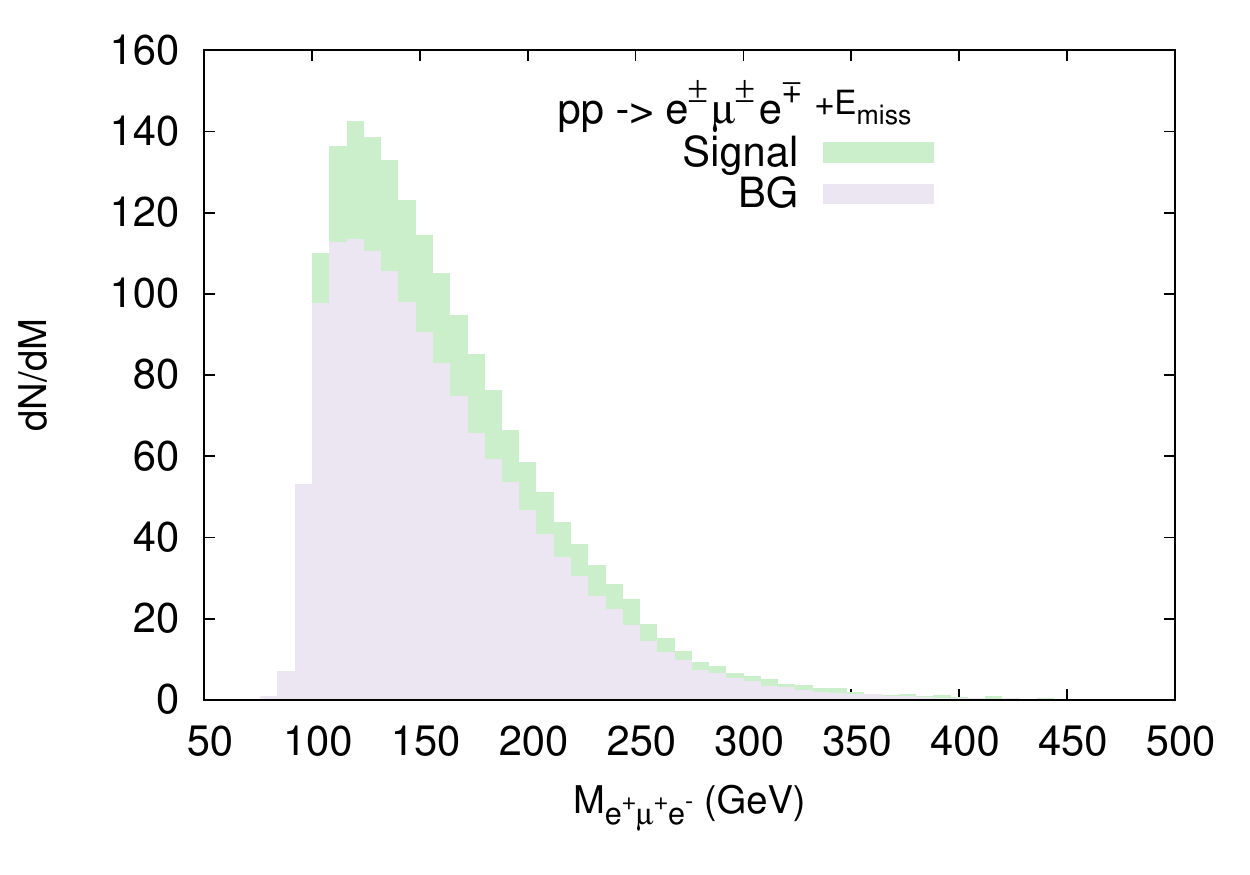}~\includegraphics [width=0.33\textwidth]{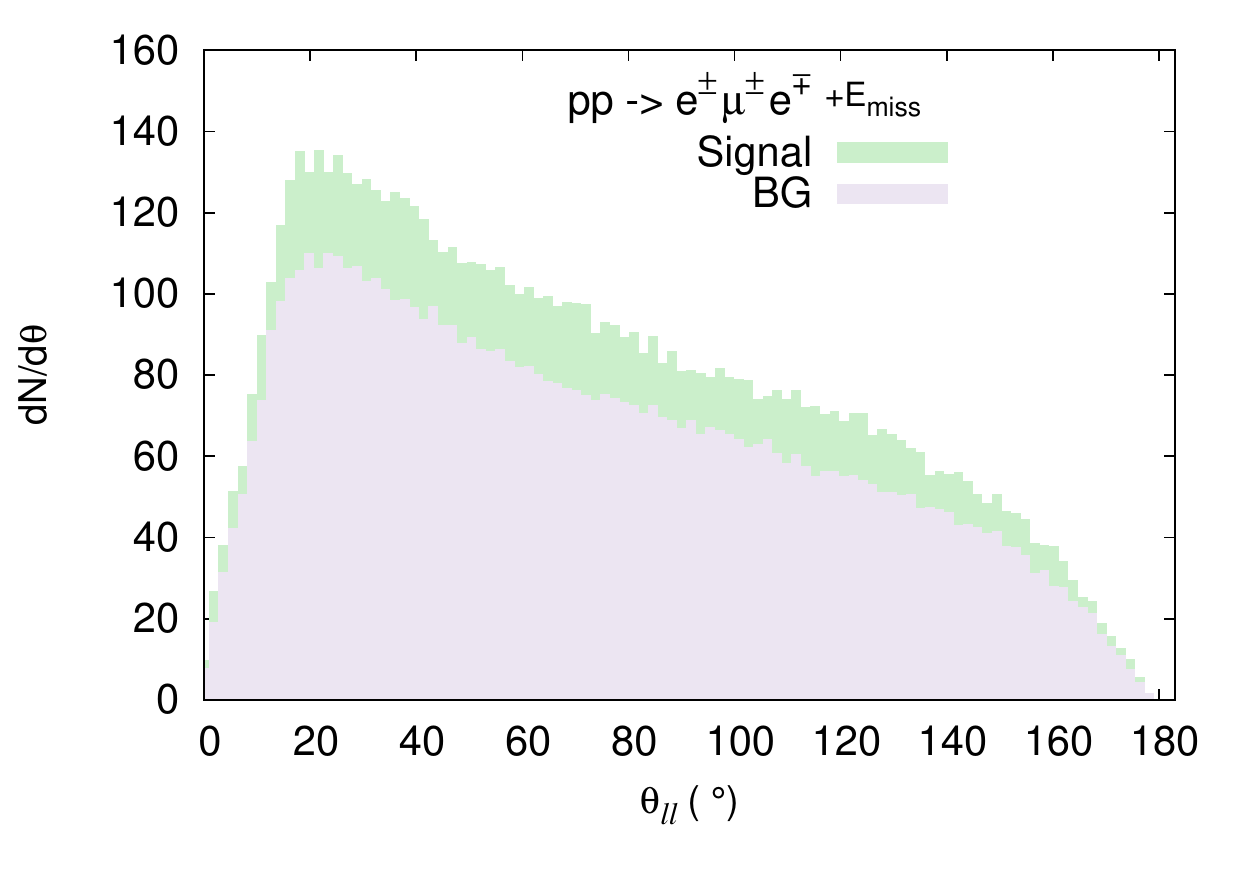}
\includegraphics [width=0.33\textwidth]{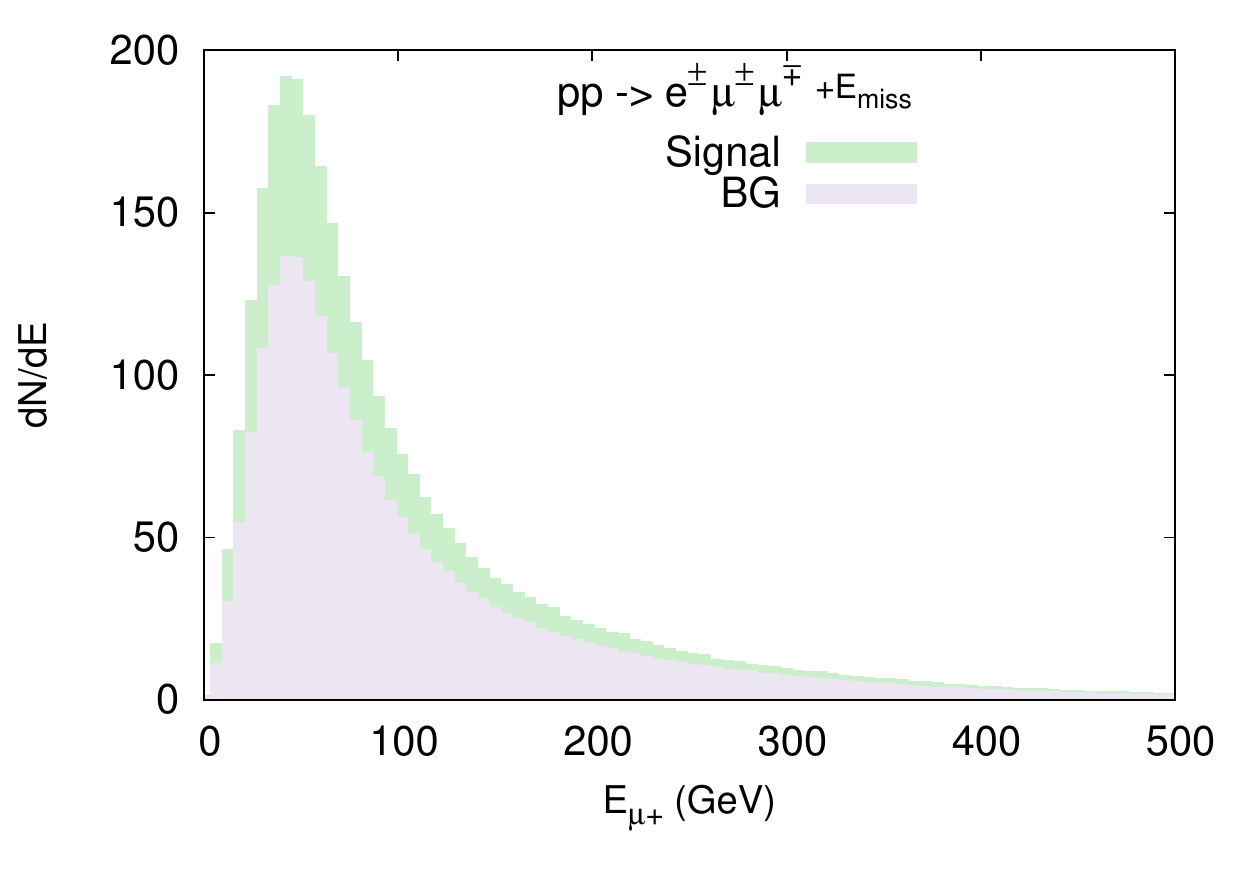}~\includegraphics [width=0.33\textwidth]{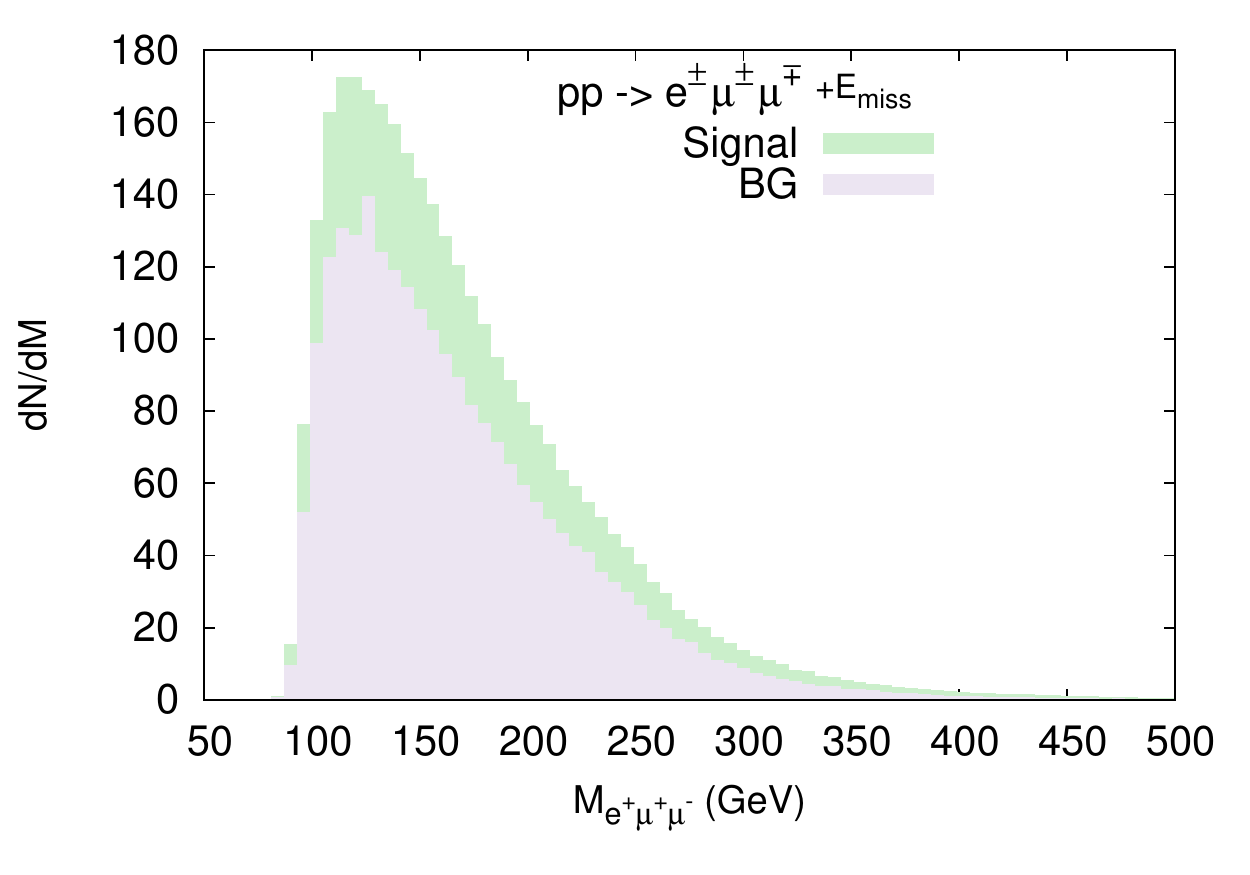}~\includegraphics [width=0.33\textwidth]{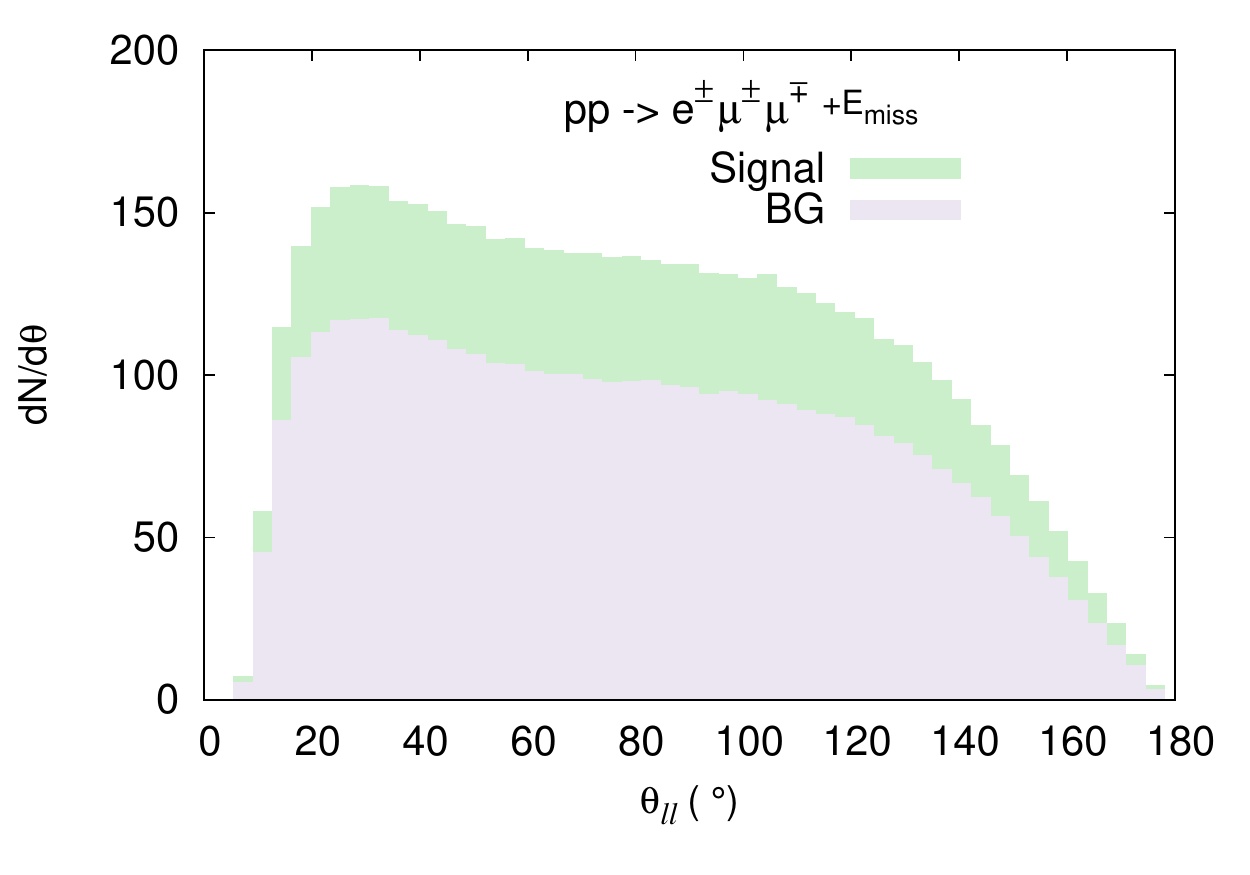} 
\par\end{centering}
\caption{Number of events of the energy distribution $E_{\ell}$, the invariant
mass distribution of the three leptons $M_{\ell\ell\ell}$, and the
angular distribution between pairs of leptons $\theta_{\ell\ell}$
at $\sqrt{s}$ = 14 TeV and $\int\mathcal{L}dt=300$ $fb^{-1}$.}
\label{kinematic-newcuts} 
\end{figure}

We note that the imposed cut values on the kinematic variables are
different than those provided by CMS, except for the range of the
invariant mass of two charged leptons $M_{\ell^{+}\ell^{-}}$, and
the pseudo rapidity $\eta^{\ell}$\,\ which still relevant for discriminating
the signal from background. Moreover, we attempt to introduce supplementary
criteria by applying cuts on the invariant masses $M_{e^{+}\mu^{+}}$
and $M_{\ell,\nu}$ of the fermion pairs $(e^{+}\mu^{+})$ and $(\ell,\nu)$,
respectively. These extra cuts allowed us to optimize the total cross
section for the signal at $\sqrt{s}=8$ TeV and $14$ TeV. This is
illustrated in Fig.~\ref{kinematic-newcuts} where we present the
angular distribution between pairs of leptons, the energy distribution
of lepton, and the invariant mass distribution of the three leptons
at $\sqrt{s}=14$ TeV. 

The kinematical distributions in Fig.~\ref{kinematic-newcuts} show
a significant excess of events which is an indication of a trilepton
signal. Clearly, there is a larger excess in the channel $e\mu\mu$
than in the $ee\mu$ channel for this benchmark. According to the
cross section values in Fig. \ref{CS}, we expect the same difference
for other benchmarks. The overall shape of the distributions for the
signal and the background looks very similar due to two reasons: (1)
the source of the event excess is the interference contribution $\sigma_{interference}$,
and (2) the cuts are chosen such that the difference $d\left(\sigma_{M}-\sigma_{BG}\right)/dX$
is strictly positive, where X represents the kinematic variables in
Tab. \ref{tab-cuts}. In Tab.~\ref{cs}, we present the cross section
values of the signal and background after imposing the cuts for the
CM energies 8 TeV and 14 TeV. The corresponding significance for each
benchmark point is shown in Tab.~\ref{tab-sign}.

\begin{table} [h]
\begin{centering}
\begin{adjustbox}{max width=\textwidth} %
\begin{tabular}{|c|cc|c|}
\hline 
Process & $\begin{array}{ccc}
B_{1}@8~TeV & & B_{2}@8~TeV\end{array}$ & & $\begin{array}{ccc}
B_{1}@14~TeV & & B_{2}@14~TeV\end{array}$\tabularnewline
\hline 
\hline 
$\sigma_{BG}\left(e^{\pm}\mu^{\pm}e^{\mp}+\not E_{T}\right)$ & 22.79 & & 40.84\tabularnewline
\hline 
$\sigma_{BG}\left(e^{\pm}\mu^{\pm}\mu^{\mp}+\not E_{T}\right)$ & 20.74 & & 46.44\tabularnewline
\hline 
\hline 
$\sigma_{EX}\left(e^{\pm}\mu^{\pm}e^{\mp}+\not E_{T}\right)$ & $\begin{array}{ccc}
28.12 & & 28.06\end{array}$ & & $\begin{array}{ccc}
49.70 & & 48.55\end{array}$\tabularnewline
\hline 
$\sigma_{EX}\left(e^{\pm}\mu^{\pm}\mu^{\mp}+\not E_{T}\right)$ & $\begin{array}{ccc}
26.13 & & 26.06\end{array}$ & & $\begin{array}{ccc}
57.28 & & 56.80\end{array}$\tabularnewline
\hline 
\end{tabular}\end{adjustbox} 
\par\end{centering}

\caption{The expected and background cross section values (in fb) at 8 TeVand
14 TeV for the two benchmark points $B_{1}$ and $B_{2}$.}
\label{cs} 
\end{table}

\begin{table} [h]
\centering\begin{adjustbox}{max width=\textwidth} %
\begin{tabular}{|c|c||c|c||c|c|}
\hline 
Process & Benchmark & $N_{20.3}$ & $S_{20.3}$ & $N_{300}$ & $S_{300}$\tabularnewline
\hline 
$pp\rightarrow e^{\pm}\mu^{\pm}e^{\mp}+\slashed{E}_{T}$ & $B_{1}$ & 108.20 & 4.53 & 2058 & 21.77\tabularnewline
\hline 
 & $B_{2}$ & 106.98 & 4.48 & 2313 & 19.16\tabularnewline
\hline 
\hline 
$pp\rightarrow e^{\pm}\mu^{\pm}\mu^{\mp}+\slashed{E}_{T}$ & $B_{1}$ & 109.42 & 4.75 & 3252 & 24.81\tabularnewline
\hline 
 & $B_{2}$ & 108.02 & 4.69 & 3108 & 23.81\tabularnewline
\hline 
\end{tabular}\end{adjustbox} \caption{The significance corresponding to the integrated luminosity values
$\mathcal{L}_{int}$ = 20.3 (300) $fb^{-1}$ at 8 TeV (14 TeV) for
the benchmark points $B_{1}$ and $B_{2}$.}
\label{tab-sign} 
\end{table}

\begin{figure} [h]
\begin{centering}
\includegraphics [width=0.5\textwidth]{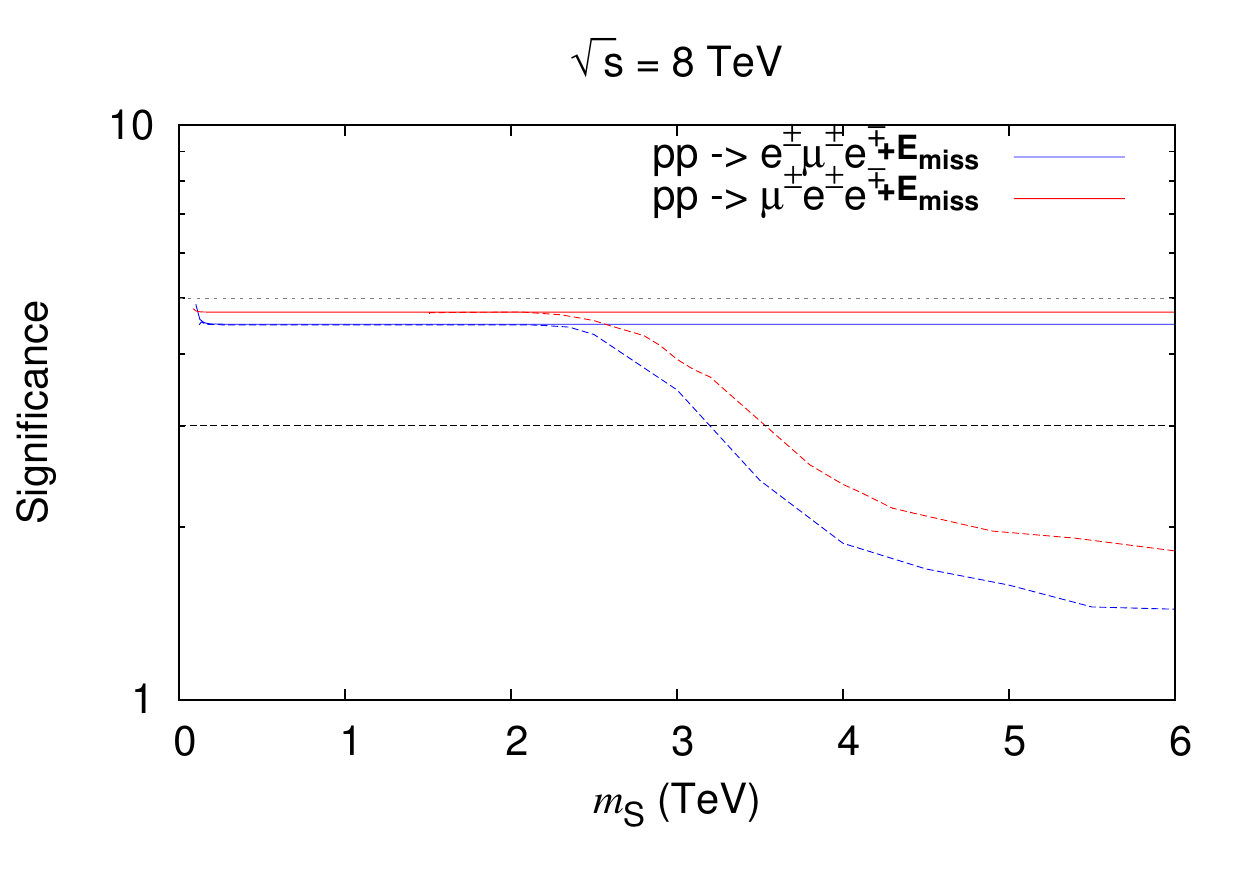}~\includegraphics [width=0.5\textwidth]{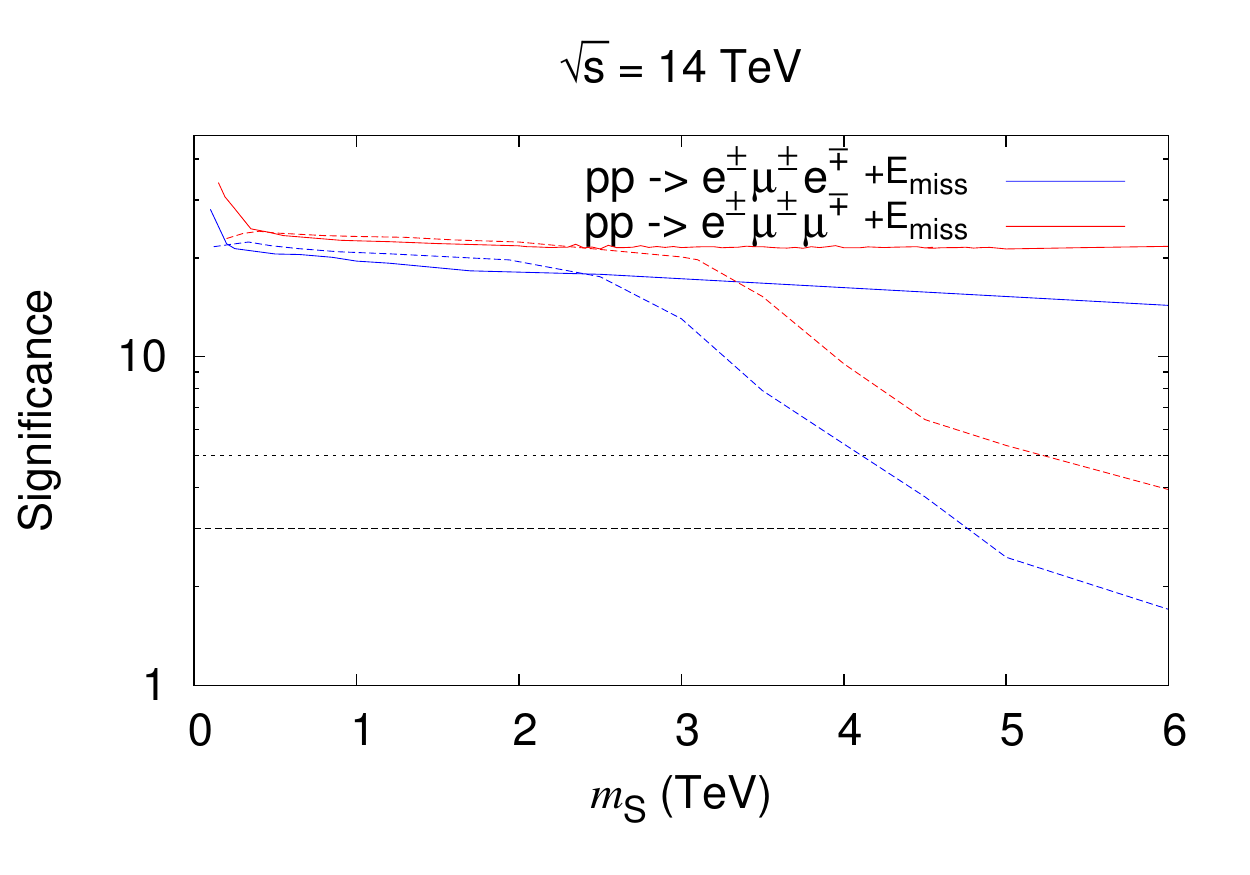} 
\par\end{centering}
\caption{Significance for the relevant process $pp\rightarrow\ell^{\pm}\ell^{\pm}\ell^{\mp}+\slashed{E}_{T}$
at $\sqrt{s}$ = 8 TeV (left) and $\sqrt{s}$ = 14 TeV (right) within
the new cuts. The black dashed horizontal lines represent the significance
value $S=$ 3, 5, respectively. solid and the dashed lines are explained
in the text.}
\label{s-newcuts} 
\end{figure}

In order to see how does the significance change with large charged
scalar mass values, we consider the benchmark point $B_{1}$ given
in Tab. \ref{tab-point}, and increase $m_{S}$ at both CM energies
8 TeV and 14 TeV for the integrated luminosity 20.3 $fb^{-1}$ and
300 $fb^{-1}$, respectively. We first keep the couplings $f_{\alpha\beta}$
to be constant and therefore the LFV constraints get relaxed with
larger $m_{S}$ values. In the second case, we vary $m_{S}$ values
while keeping LFV observables, such as $B(\ell_{\alpha}~\rightarrow\ell_{\beta}+\gamma)$,
constant. The two cases are shown in Fig.~\ref{s-newcuts} with dashed
and solid lines, respectively. Thus, whatever the values of charged
scalar mass or the LFV branching ratios, the significance should lie
in between these two curves. We can see from the figure that the significance
can reach 3$\sigma$ for any charged scalar $S^{\pm}$ mass under
2 TeV, and $5\sigma$ is ensured until $m_{S}=3$ TeV in the case
where $\sqrt{s}=14$ TeV.

We remark here that the Feynman diagrams that mediate the processes
$pp\rightarrow\ell^{\pm}\ell^{\pm}\ell^{\mp}+\slashed{E}_{T}$ can
be classified as SM and non-SM diagrams with amplitudes $\mathcal{M}_{SM}$
and $\mathcal{M}_{S}$, respectively. Therefore, the event number
difference $N_{ex}=N_{M}-N_{BG}$ is proportional to the combination
$\sigma_{interference}\varpropto Re\left(\mathcal{M}_{SM}^{\dag}\mathcal{M}_{S}\right)$,
since $\sigma_{S}$ is negligible as mentioned previously. In other
words, the significance shown in Fig. \ref{s-newcuts} is directly
proportional to the couplings combination $\left\vert f_{\alpha\rho}f_{\beta\rho}\right\vert ^{2}$
that appears in the expressions of the branching ratios of the processes
$\mu\rightarrow e+\gamma$ and $\tau\rightarrow\mu+\gamma$. This
means that there is a direct correlation between the discovery of
the LFV processes and the signals.

In our analysis at $\sqrt{s}=14$ TeV, we have presented the points
which can be discovered with an integrated luminosity of 300 $fb^{-1}$.
However, another way to probe the interaction (\ref{LL}) is to extend
our analysis by considering a maximally LFV process like the process
$pp\rightarrow e^{\pm}\mu^{\pm}\tau^{\mp}+\slashed{E}_{T}$, where
the tau lepton can be identified through its hadronic decay~\cite{Br-tau}
rather than its leptonic one in order to avoid an additional source
of missing energy. In addition to the case of $\sqrt{s}=14$ TeV,
we consider also the very high energy such as at the HL-LHC $\sqrt{s}=100$
TeV. Then, the event number here is given by 
\begin{equation}
N_{e\mu\tau}=L\times\sigma(pp\rightarrow e^{\pm}\mu^{\pm}\tau^{\mp}+\slashed{E}_{T})\mathcal{B}(\tau\rightarrow hadrons),\label{Nev}
\end{equation}
where the corresponding background event number is given by 
\begin{equation}
N_{BG}=L\times\sigma(pp\rightarrow WWW)\mathcal{B}(W\rightarrow e\nu)\mathcal{B}(W\rightarrow\mu\nu)\mathcal{B}(W\rightarrow\tau\nu)\mathcal{B}(\tau\rightarrow hadrons).\label{BG}
\end{equation}

We find that the significance is so small at both $\sqrt{s}$ =14
TeV and 100 TeV for luminosity values of the order $\mathcal{O}(\text{ab}^{-1})$.
Detailed investigation is required to reach final conclusion about
the possibility of detecting the maximally LFV in this model (\ref{LL}).

\section{Summary}

In this paper, we investigate the effect of a singlet charged scalar
at the LHC by performing a detailed analysis of three isolated leptons
in the final state. First we applied the same cuts used by the CMS
collaboration at 8 TeV on a large number of benchmarks that are consistent
with LFV bounds and we found no significant deviation from the SM.
Whereas, within the same cuts we expect significant deviation at 14
TeV. So to enhance the signal over the background, we applied new
cuts for both 8 TeV and 14 TeV. We have chosen two benchmark points
$B_{1}$ and $B_{2}$ with different values of $m_{S}$ in order to
probe the effect of this charged scalar in the tripleton channel and
we found that a deviation from the SM can be seen using $8~$TeV data
and expect that a discovery is potentially possible at $14~$TeV.
Using our analysis of 8 TeV (14 TeV), we can exclude charged scalar
masses $m_{S}<~3$ TeV ($m_{S}<4~\mathrm{TeV}$). We found that the
significance is directly proportional to $\mathcal{B}(\ell_{\alpha}~\rightarrow\ell_{\beta}+\gamma)$,
and hence there is a direct correlation between the LFV discovery
and our signal.

Another way to search for the trilepton signal is via the maximally
LFV processes such $e^{\pm}\mu^{\pm}\tau^{\mp}$, where the tau lepton
is identified through its hadronic decay. However, even at $\sqrt{s}=100$
TeV the significance is too small for luminosity values of the order
$\mathcal{O}(ab^{-1})$.

\subsection*{Acknowledgments}

We would like to thank C.-S. Chen and J.M. No for useful discussion;
and A.B. Hammou for valuable comments on the manuscript. C.G. and
D.C. would like to thank the ICTP for the warm hospitality during
part of this work. A.A. is supported by the Algerian Ministry of Higher
Education and Scientific Research under the CNEPRU Project No B00L02UN180120140040.


\begin{thebibliography}{10}
\bibitem{seesaw}P. Minkowski, Phys. Lett. B 67, 421 (1977); M. Gell-Mann,
P. Ramond and R. Slansky, Proceedings of the Supergravity Stony Brook
Workshop, New York 1979, eds. P. Van Nieuwenhuizen and D. Freedman;
T. Yanagida, Proceedings of the Workshop on Unified Theories and Baryon
Number in the Universe, Tsukuba, Japan 1979, eds. A. Sawada and A.
Sugamoto; R. N. Mohapatra and G. Senjanovic, Phys. Rev. Lett. 44,
912 (1980).

\bibitem{seesaw0}J. Schechter and J.W.F. Valle, Phys. Rev. D 22,
2227 (1980).

\bibitem{seesawII}T.~P.~Cheng and L.~F.~Li, 
Phys.\ Rev.\ D\textbf{22} (1980) 2860; M.~Magg and C.~Wetterich, 
Phys.\ Lett.\ B\textbf{94} (1980) 61; C.~Wetterich, 
Nucl.\ Phys.\ B\textbf{187} (1981) 343; R.~N.~Mohapatra and G.~Senjanovic, 
Phys.\ Rev.\ D\textbf{23} (1981) 165.

\bibitem{seesawIII}R.~Foot, H.~Lew, X.~G.~He and G.~C.~Joshi,
Z.\ Phys.\ C\textbf{44} (1989) 441.

\bibitem{DHT}A.~de Gouvea, D.~Hernandez and T.~M.~P.~Tait, 
Phys.\ Rev.\ D\textbf{89}, no. 11, 115005 (2014).

\bibitem{Boucenna:2014zba}S.~M.~Boucenna, S.~Morisi and J.~W.~F.~Valle,
Adv.\ High Energy Phys.\ \textbf{2014}, 831598 (2014) 
 [arXiv:1404.3751 [hep-ph]]; F.~Borzumati and Y.~Nomura, 
Phys.\ Rev.\ D\textbf{64} (2001) 053005 
 [hep-ph/0007018]; M.~Fabbrichesi and S.~T.~Petcov, 
Eur.\ Phys.\ J.\ C\textbf{74} (2014) 2774 
 [arXiv:1304.4001 [hep-ph]]; P.~H.~Gu, M.~Hirsch, U.~Sarkar and J.~W.~F.~Valle,
Phys.\ Rev.\ D\textbf{79} (2009) 033010 
 [arXiv:0811.0953 [hep-ph]]; P.~Fileviez Perez, T.~Han and T.~Li, 
Phys.\ Rev.\ D\textbf{80} (2009) 073015 
 [arXiv:0907.4186 [hep-ph]]; E.~Ma, 
Mod.\ Phys.\ Lett.\ A\textbf{24} (2009) 2491 
 [arXiv:0905.2972 [hep-ph]]; A.~Ahriche, S.~M.~Boucenna and S.~Nasri, 
Phys.\ Rev.\ D\textbf{93} (2016) no.7, 075036 
 [arXiv:1601.04336 [hep-ph]].

\bibitem{rad}A. Zee, Phys. Lett. B 161, 141 (1985); A. Zee, Nucl.
Phys. B 264, 99 (1986); K.S. Babu, Phys. Lett. B 203, 132 (1988).

\bibitem{Ma}E.~Ma, 
Phys.\ Rev.\ Lett.\ \textbf{81}(1998) 1171 
 [hep-ph/9805219].

\bibitem{knt}L.~M.~Krauss, S.~Nasri and M.~Trodden, 
Phys.\ Rev.\ D\textbf{67}, 085002 (2003) [hep-ph/0210389].

\bibitem{kanemura}M. Aoki, S. Kanemura and O. Seto, Phys. Rev. Lett.
102, 051805 (2009) [arXiv:0807.0361 [hep-ph]]; M. Aoki, S.
Kanemura and O. Seto, Phys. Rev. D 80 (2009) 033007 [arXiv:0904.3829
 [hep-ph]].

\bibitem{Okada:2016rav}H.~Okada and K.~Yagyu, 
Phys.\ Lett.\ B\textbf{756} (2016) 337 
 [arXiv:1601.05038 [hep-ph]]. 
Y.~Farzan, 
JHEP \textbf{1505} (2015) 029 
 [arXiv:1412.6283 [hep-ph]]. 
D.~Aristizabal Sierra, A.~Degee, L.~Dorame and M.~Hirsch, 
JHEP \textbf{1503} (2015) 040 
 [arXiv:1411.7038 [hep-ph]]. 
Y.~Farzan, S.~Pascoli and M.~A.~Schmidt, 
JHEP \textbf{1303} (2013) 107 
 [arXiv:1208.2732 [hep-ph]]. 
Y.~Farzan and E.~Ma, 
Phys.\ Rev.\ D \textbf{86} (2012) 033007 
 [arXiv:1204.4890 [hep-ph]]. 
Y.~Farzan, S.~Pascoli and M.~A.~Schmidt, 
JHEP \textbf{1010} (2010) 111 
 [arXiv:1005.5323 [hep-ph]]. 
Y.~Farzan, 
Phys.\ Rev.\ D \textbf{80} (2009) 073009 
 [arXiv:0908.3729 [hep-ph]]. 
C.~Boehm, Y.~Farzan, T.~Hambye, S.~Palomares-Ruiz and S.~Pascoli,
Phys.\ Rev.\ D \textbf{77} (2008) 043516 
 [hep-ph/0612228]. 
D.~Restrepo, O.~Zapata and C.~E.~Yaguna, 
JHEP \textbf{1311} (2013) 011 
 [arXiv:1308.3655 [hep-ph]]. 
P.~W.~Angel, Y.~Cai, N.~L.~Rodd, M.~A.~Schmidt and R.~R.~Volkas,
JHEP \textbf{1310} (2013) 118 Erratum: [JHEP \textbf{1411} (2014)
092] 
 [arXiv:1308.0463 [hep-ph]]. 

\bibitem{Cheung}K.~Cheung and O.~Seto, 
Phys.\ Rev.\ D\textbf{69}, 113009 (2004) [hep-ph/0403003]; A.~Ahriche, K.~L.~McDonald
and S.~Nasri, Phys.\ Rev.\ D \textbf{92}, no. 9, 095020 (2015)
 [arXiv:1508.05881].

\bibitem{AN}A.~Ahriche and S.~Nasri, 
JCAP\textbf{1307}, 035 (2013) [arXiv:1304.2055]; E.~A.~Baltz and
L.~Bergstrom, 
Phys.\ Rev.\ D\textbf{67}, 043516 (2003) [hep-ph/0211325].

\bibitem{knt3}A.~Ahriche, C.~S.~Chen, K.~L.~McDonald and S.~Nasri,
Phys.\ Rev.\ D\textbf{90}, 015024 (2014) [arXiv:1404.2696 [hep-ph]]; 
A.~Ahriche, K.~L.~McDonald and S.~Nasri, 
JHEP\textbf{1410}, 167 (2014) [arXiv:1404.5917 [hep-ph]]; 
A.~Ahriche, K.~L.~McDonald, S.~Nasri and T.~Toma, 
Phys.\ Lett.\ B\textbf{746}, 430 (2015) [arXiv:1504.05755 [hep-ph]].

\bibitem{knt-SI}A.~Ahriche, K.~L.~McDonald and S.~Nasri, 
JHEP\textbf{1602}, 038 (2016) [arXiv:1404.5917 [hep-ph]].

\bibitem{Adam:2013mnn}J.~Adam \textit{et al.} [MEG Collaboration],
Phys.\ Rev.\ Lett.\textbf{110} (2013) 201801 
 [arXiv:1303.0754 [hep-ex]].

\bibitem{PDG}K.~A.~Olive \textit{et al.} [Particle Data Group
Collaboration], ``Review of Particle Physics,''Chin.\ Phys.\ C\textbf{38}, 090001 (2014).

\bibitem{Perez:2008zc}W.~Y.~Keung and G.~Senjanovic, 
Phys.\ Rev.\ Lett.\ \textbf{50}, 1427 (1983); P.~Fileviez Perez, T.~Han, G.~Y.~Huang, T.~Li and
K.~Wang, 
Phys.\ Rev.\ D\textbf{78}, 071301 (2008) [arXiv:0803.3450 [hep-ph]]; 
S.~Gabriel, B.~Mukhopadhyaya, S.~Nandi and S.~K.~Rai, 
Phys.\ Lett.\ B\textbf{669}, 180 (2008) [arXiv:0804.1112 [hep-ph]]; 
C.~S.~Chen, C.~Q.~Geng, J.~N.~Ng and J.~M.~S.~Wu, 
JHEP\textbf{0708}, 022 (2007) [arXiv:0706.1964 [hep-ph]]; 
J.~Kersten and A.~Y.~Smirnov, 
Phys.\ Rev.\ D\textbf{76}, 073005 (2007) [arXiv:0705.3221 [hep-ph]]; 
A.~Das and N.~Okada, 
Phys.\ Rev.\ D\textbf{88}, 113001 (2013) [arXiv:1207.3734 [hep-ph]]; D.~Atwood,
S.~Bar-Shalom and A.~Soni, 
Phys.\ Rev.\ D\textbf{76}, 033004 (2007) [hep-ph/0701005]; 
S.~Kanemura, T.~Nabeshima and H.~Sugiyama, 
Phys.\ Rev.\ D\textbf{87}, no. 1, 015009 (2013) [arXiv:1207.7061 [hep-ph]];
P.~S.~B.~Dev and A.~Pilaftsis, 
Phys.\ Rev.\ D\textbf{86}, 113001 (2012) 
 [arXiv:1209.4051 [hep-ph]]; S.~Antusch and O.~Fischer, 
JHEP \textbf{1410}, 094 (2014) 
 [arXiv:1407.6607 [hep-ph]]; S.~Antusch, E.~Cazzato and O.~Fischer, 
JHEP\textbf{1604}, 189 (2016) 
 [arXiv:1512.06035 [hep-ph]]; S.~Antusch, E.~Cazzato and O.~Fischer, 
arXiv:1604.02420 [hep-ph].

\bibitem{knt-ilc}A.~Ahriche, S.~Nasri and R.~Soualah, 
Phys.\ Rev.\ D\textbf{89}, 095010 (2014) [arXiv:1403.5694 [hep-ph]].

\bibitem{Guella}C.~Guella, D.~Cherigui, A.~Ahriche, S.~Nasri and R.~Soualah,
 Phys.\ Rev.\ D {\bf 93} (2016) no.9, 095022
 [arXiv:1601.04342 [hep-ph]]; A.~Pilaftsis, 
 Z.\ Phys.\ C\textbf{55}, 275 (1992) 
 [hep-ph/9901206].

\bibitem{Tri}A.~Das, N.~Nagata and N.~Okada, 
JHEP\textbf{1603}, 049 (2016) 
 [arXiv:1601.05079 [hep-ph]]; A.~Chatterjee, N.~Chakrabarty and B.~Mukhopadhyaya,
Phys.\ Lett.\ B\textbf{754}, 14 (2016).

\bibitem{CMS-cuts}S.~Chatrchyan {\it et al.} [CMS Collaboration],
 Phys.\ Rev.\ D {\bf 90} (2014) 032006
 [arXiv:1404.5801 [hep-ex]]. For an earlier analysis with $\sqrt{s}$ = 7 TeV LHC data, see S.
Chatrchyan et al. [CMS Collaboration], JHEP 1206, 169 (2012) [arXiv:1204.5341
 [hep-ex]].

\bibitem{Zee-ref}For example: K.~S.~Babu and C.~Macesanu, 
Phys.\ Rev.\ D\textbf{67} (2003) 073010 
 [hep-ph/0212058].; S.~Kanemura, T.~Nabeshima and H.~Sugiyama, 
Phys.\ Lett.\ B\textbf{703} (2011) 66 
 [arXiv:1106.2480 [hep-ph]]; M.~Lindner, D.~Schmidt and T.~Schwetz, 
Phys.\ Lett.\ B\textbf{705} (2011) 324 
 [arXiv:1105.4626 [hep-ph]]; S.~S.~C.~Law and K.~L.~McDonald, 
Int.\ J.\ Mod.\ Phys.\ A\textbf{29} (2014) 1450064 
 [arXiv:1303.6384 [hep-ph]]; C.~S.~Chen, C.~Q.~Geng and D.~V.~Zhuridov,
arXiv:0806.2698 [hep-ph]; K.~Nishiwaki, H.~Okada and Y.~Orikasa, 
Phys.\ Rev.\ D\textbf{92} (2015) no.9, 093013 
 [arXiv:1507.02412 [hep-ph]]; T.~Nomura, H.~Okada and Y.~Orikasa, 
Phys.\ Rev.\ D\textbf{93} (2016) no.11, 113008 
 [arXiv:1603.04631 [hep-ph]]; T.~Nomura, H.~Okada and Y.~Orikasa, 
arXiv:1602.08302 [hep-ph].

\bibitem{Belyaev:2012qa}A.~Belyaev, N.~D.~Christensen and A.~Pukhov,
Comput.\ Phys.\ Commun.\ \textbf{184}, 1729 (2013) [arXiv:1207.6082 [hep-ph]].

\bibitem{CMS:2010lua} [CMS Collaboration], 
CMS-PAS-MUO-10-002.

\bibitem{Chatrchyan:2012ya}S.~Chatrchyan \textit{et al.} [CMS
Collaboration], 
Eur.\ Phys.\ J.\ C\textbf{72}, 2189 (2012) 
 [arXiv:1207.2666 [hep-ex]].

\bibitem{CMS:2010bta} [CMS Collaboration], 
CMS-PAS-EGM-10-004.

\bibitem{okada}A.~Das, P.~S.~Bhupal Dev and N.~Okada, 
Phys.\ Lett.\ B\textbf{735}, 364 (2014) [arXiv:1405.0177 [hep-ph]]; A.~Das
and N.~Okada, 
Phys.\ Rev.\ D\textbf{93} (2016) 033003 
 [arXiv:1510.04790 [hep-ph]].

\bibitem{Br-tau}A.~J.~Weinstein, 
Nucl.\ Phys.\ Proc.\ Suppl.\textbf{76}, 497 (1999) [hep-ex/9811044].

\end{thebibliography}
\end{document}